\documentclass[fleqn,final,3p,11pt]{elsarticle}

\usepackage{amsmath,amssymb,amsfonts}
\usepackage{amsthm}
\usepackage{multirow}
\usepackage{rotating}
\usepackage{array,multirow,graphicx}
\usepackage{color, colortbl}
\usepackage{tabularx}
\usepackage{caption}
\usepackage{multirow}
\usepackage{bbm}
\usepackage[outdir=./]{epstopdf}

\usepackage{afterpage}
\usepackage{booktabs}
\usepackage{graphicx}
\usepackage{placeins,slashbox}
\usepackage{amssymb}
\usepackage{amsmath,amsthm,amssymb,amsfonts}
\usepackage{amsfonts}
\usepackage{graphicx}
\usepackage{subcaption}

\newtheorem{Theorem}{Theorem}[section]

\newtheorem{Proposition}[Theorem]{Proposition}

 \numberwithin{equation}{section}

\UseRawInputEncoding

\begin{document}

\begin{frontmatter}


\title{Value-at-Risk-Based Portfolio Insurance: Performance Evaluation and Benchmarking Against CPPI in a Markov-Modulated Regime-Switching Market}

\author[add1]{Peyman Alipour}
\ead{palipour@stevens.edu}
\author[add2]{Ali Foroush Bastani}
\ead{bastani@iasbs.ac.ir}

\address[add1]{School of Business, Stevens Institute of Technology, \\ P.O. Box 07030, Hoboken, NJ, USA}
\address[add2]{Department of Mathematics, Institute for Advanced Studies in Basic Sciences,\\
P.O. Box 45195-1159,  Zanjan, Iran}



\begin{abstract}
Designing dynamic portfolio insurance strategies under market conditions switching between two or more regimes is a challenging task in financial economics. Recently, a promising approach employing the value-at-risk (VaR) measure to assign weights to risky and riskless assets has been proposed in [Jiang C., Ma Y. and An Y. ``The effectiveness of the VaR-based portfolio insurance strategy: An empirical analysis'', {\rm International Review of Financial Analysis} 18(4) (2009): 185-197]. In their study, the risky asset follows a geometric Brownian motion with constant drift and diffusion coefficients. In this paper, we first extend their idea to a regime-switching framework in which the expected return of the risky asset and its volatility depend on an unobservable Markovian term which describes the cyclical nature of asset returns in modern financial markets. We then analyze and compare the resulting VaR-based portfolio insurance (VBPI) strategy with the well-known constant proportion portfolio insurance (CPPI) strategy. In this respect, we employ a variety of performance evaluation criteria such as Sharpe, Omega and Kappa ratios to compare the two methods. Our results indicate that the CPPI strategy has a better risk-return tradeoff in most of the scenarios analyzed and maintains a relatively stable return profile for the resulting portfolio at the maturity.
\end{abstract}
\begin{keyword}
Finance, Constant Proportion Portfolio Insurance, Value at Risk, Regime-Switching, Omega Performance Measure.
\vspace{0.3cm}
\end{keyword}
\end{frontmatter}

%

\section{Introduction}
Portfolio insurance products are popular structured investment tools widely used by private and institutional investors, providing their holders with capital protection in down turning markets, while allowing benefits from upside market potentials (see e.g. \cite{lee2012security,luskin1988portfolio,prigent2007portfolio} and the many references therein). The first milestone research in this direction is due to Leland and Rubinstein \cite{leland1988evolution} in 1976 who proposed a synthetically constructed
put option by trading on the underlying risky portfolio and a risk-free asset to dynamically hedge the risk of  issuer/guarantor liability. At the same time, Brennan and Schwartz \cite{brennan1976pricing}, studying equity-linked life insurance policies guaranteeing a minimum return, reached a similar result by using the then new arbitrage-free pricing methodology of Black and Scholes \cite{black1973pricing} and Merton \cite{merton1976option}. These ideas were culminated in option-based portfolio insurance (OBPI) strategies which invest the initial endowment in a risky reference portfolio covered by a put option with a strike chosen to be proportional to the guaranteed amount \cite{basak2002comparative}.

In a parallel line of research, Perold \cite{perold1986constant} (see also Perold and Sharpe \cite{perold1988dynamic} and Black and Jones \cite{black1987simplifying}) introduced an alternative dynamic trading strategy, called constant proportion portfolio insurance (CPPI), based on continuous rebalancing of a portfolio containing a safe asset (e.g. treasury bills) and a risky one (e.g. a financial index) in response to fluctuations in market conditions. By choosing a floor value, $F_T$, as the minimum acceptable portfolio level at the maturity, $T$, at each rebalancing time, $t$, the difference between the portfolio value, $V_t$ and the (discounted) floor is computed as $C_t=V_t-F_Te^{-r(T-t)}$ (called the {\it cushion}) and the {\it exposure} to the risky asset is determined by $E_t=mC_t$ in which $m$ is called the {\it multiple}. The excess will naturally be invested in the riskless asset.

It is shown (see e.g. Black and Perold \cite{black1992theory}) that in the absence of borrowing constraints and transaction costs, CPPI is a special case of the HARA utility-maximizing rules that have appeared in the literature of continuous-time asset allocation (see e.g. Merton \cite{merton1975optimum}). For more details on the basic CPPI model and its extensions, see e.g. Hirsa \cite{hirsa2010constant}, Boulier and Kanniganiti \cite{boulier2005expected}, Temocin, et al. \cite{temocin2018constant}, Bertrand and Prigent \cite{bertrand2016portfolio}, Ben Ameur and Prigent \cite{ameur2014portfolio,ameur2018risk} among many other references.

When the basic CPPI strategy is exploited with a constant multiple during the holding-period, some inevitable shortcomings will arise due mainly to ignoring the investor's beliefs and risk preferences (see e.g. Hakanoglu et al. \cite{hakanoglu1989constant}). This drawback has led the researchers to the family of dynamic proportion portfolio insurance (DPPI) strategies which basically adopt the choice of the multiple (and so the risk exposure) to the volatility of market prices (see e.g. Chen and Chang \cite{chen2005dynamical}, Chen et al. \cite{chen2008dynamic}, Hamidi et al. \cite{hamidi2009risk}). The basic idea is to vary the multiple in such a way that some ``guarantee condition'' at the maturity is satisfied with a given probability for specified market conditions. This condition is closely related to the probability of violating the floor protection widely known in the literature as the {\it gap risk} (see e.g. Jessen \cite{ameur2014portfolio,jessen2014constant}).

Value-at-Risk (VaR) based portfolio insurance (VBPI) is a new approach to dynamic asset allocation which is based on the portfolio's VaR concept (see e.g. \cite{basak2001value} for a general overview of VaR-based risk management strategies). For a given time horizon, $t$ and confidence level, $p$, the value at risk is the loss in market value over the time horizon $t$ that is exceeded with probability $1-p$. In this strategy, the insured portfolio is constructed and rebalanced frequently in such a way that the portfolio level at each time step exceeds the floor at a given confidence level \cite{jiang2009effectiveness}. Based on the fact that this approach targets the gap risk faced by the insurer, it leads naturally to a path-dependent portfolio strategy consistent with the performance measure used to evaluate the insurance strategy. It could also be viewed as a generalized CPPI strategy which provides the buyer with additional flexibility to benefit more from upward market movements while limiting the potential loss from downward moves.

While the basic CPPI strategy has been proposed and analyzed mainly for diffusion-based dynamics of the underlying risky portfolio, there have been some efforts in the literature to extend this methodology to more complex situations in which the underlying assete follows a dynamic process such as jump-diffusion (see e.g. \cite{cont2009constant,chakrabarty2017theoretical}), regime-switching diffusion (see e.g. \cite{hainaut2011risk}) and regime-switching exponential L\'{e}vy model (see e.g. \cite{weng2013constant}). However, a systematic study in the literature is still lacking for the extension of the VaR-based DPPI strategies to such families of asset return distributions. As a first step to fill this gap, we consider here a continuous-time regime-switching market in which the model parameters switch from one regime to another according to an unobservable Markov process (see e.g. \cite{elliott2008hidden} and the many references therein). These models provide a natural way to capture discrete shifts in market behavior in an efficient and flexible way.

We compare the performance of constrained CPPI and VBPI strategies in this case to demonstrate the effect of regime shifts. In this respect, we perform a Monte-Carlo simulation study by generating sample paths from the underlying asset's risky dynamics and show that the CPPI strategy has a better risk-return tradeoff in most of the scenarios examined and maintains a relatively stable return profile for the resulting portfolio at the maturity. This complements the available results in the literature of portfolio insurance which claim the effectiveness of CPPI-based strategies in the presence of realistic circumstances and incompleteness assumptions on the market (see, e.g. \cite{pezier2013best}).

The structure of this paper is as follows: In Section 2, we provide the necessary background material about regime-switching diffusion processes. Section 3 in concerned with the details of stylized and constrained CPPI strategies when the underlying risky asset follows a regime-switchning geometric Brownian motion process. In Section 4, we first present the details of our proposed VaR-based strategy for the regime-switching case and then demonstrate the required computational procedure to estimate the VaR quantity needed in our proposed algorithm. Section 5 is devoted to introduce the risk-adjusted performance measures used in this paper and also the downside risk measures employed to evaluate the possible pitfalls of the proposed strategy. In the remainder and Section 6, we present the details of our comprehensive numerical experiments to validate and benchmark the VaR-based strategy against the well-known CPPI method. We conclude the paper by commenting on some possible research directions in this field.


\section{The Financial Market Model}
In the sequel, we consider an investment horizon of $[0,T]$ and assume that the value of the riskless asset, denoted by $B$, grows with a constant risk-free rate, $r$ as
\begin{equation}\label{dynamic of riskfree}
    \displaystyle{dB_t = rB_tdt.}
\end{equation}
On the other hand, the market value of the risky asset, denoted by $S$, is given by a regime-switching geometric Brownian motion (GBM) of the form
\begin{equation}\label{rs-gbm}
    \displaystyle{dS_t = \mu_{\alpha_t} S_t dt + \sigma_{\alpha_t} S_t dW_t,}
\end{equation}
with a positive initial value, $S_0$. In the above equation, $W = {(W_t)}_{0 \leqslant t
\leqslant T}$ is a standard Wiener process defined on a complete filtered probability space, $(\Omega, \mathcal{F}, \{\mathcal{F}_t\}_{0 \leqslant t
\leqslant T}, \mathbb{P})$ and $\mu_{\alpha_t}$ and $\sigma_{\alpha_t}$ are the drift and diffusion terms depending on a continuous-time stationary Markov process, ${(\alpha_t)}_{0 \leqslant t \leqslant T}$, independent of $W$.

The process, ${(\alpha_t)}_{0 \leqslant t \leqslant T}$, takes values in the set $\mathcal{H}=\{1,2,....H\}$ where each element represents a possible economic or financial regime or state of the world. The generator of this Markov chain is given by the matrix $Q=(q_{ij})_{H\times H}$ in which $q_{ij}\geqslant0$ for all $i\neq j$ and $\sum_{j=1}^H {q_{ij}} = 0,~i \in \mathcal{H}$. The transition probability matrix could be obtained as
\begin{equation}\label{PQ}
    \displaystyle {P(t,s) = \exp(Q(s-t)),}
\end{equation}
for each $s,t~(0< s\leq t)$ with the elements
\begin{equation}\label{pij}
    \displaystyle{p_{ij}(t,s) = {\Bbb P}(\alpha_s = j| \alpha_t = i), \quad i,j \in \mathcal{H}.}
\end{equation}
The probability of being in a specific state at time $t$ will be denoted by
$p_i(t)$ and is given by
\begin{equation}\label{pi}
    \displaystyle{p_i(t) = {\Bbb P}(\alpha_t =i) = \sum_{k=1}^H p_k(0) p_{ki}(0,t), \quad \forall i \in \mathcal{H}.}
\end{equation}
Also the stationary transition probabilities corresponding to the Markov process are given by
\begin{equation}\label{p&}
    p_i = \lim_{t\rightarrow\infty} p_i(t), \quad \forall i \in \mathcal{H}.
\end{equation}
In the special case where the continuous-time Markov chain, $\alpha_t$, contains only two states (e.g. state 1 denoting a stable low-volatility regime and state 2 denoting a more unstable high volatility regime), the matrix $Q$ could be written as
\begin{equation}\label{QQ}
    Q = \left(%
\begin{array}{cc}
  -q_{11} & q_{11} \\
  q_{22} & -q_{22} \\
\end{array}%
\right),
\end{equation}
with positive $q_{ij}$s and the stationary probabilities are obtained as
\begin{equation}\label{p1p2}
\begin{array}{cc}
  p_1=\frac{q_{11}}{q_{11}+q_{22}}, \quad & \quad p_2=\frac{q_{22}}{q_{11}+q_{22}}.
\end{array}
\end{equation}
In the remainder, we describe the details of CPPI and VBPI strategies analysed in this paper.
\section{Stylized Constant Proportion Portfolio Insurance}
CPPI is a dynamic self-financing strategy in which positions in risky and risk free assets are rebalanced dynamically so that the terminal value of the portfolio lies above a guaranteed level, $F_T$ (the floor), which is usually given as a percentage, $\pi$ ($0\leq \pi \le 1$), of the initial investment
\begin{equation}\label{FT}
    \displaystyle{F_T = \pi V_0,}
\end{equation}
and the value of the floor at any given time, $t\in [0,T]$, is obtained as
\begin{equation}\label{Ft}
    \displaystyle{F_t = e^{-r(T-t)}F_T.}
\end{equation}
The difference between the market value of the portfolio, ${V_t}^{\rm CPPI}$ and the floor is called the cushion and is denoted by $C_t$.

In the standard CPPI strategy, there is no restriction on the risky part of the portfolio\footnote{The standard CPPI strategy allows the exposure to be leveraged at any level i.e., there is no constraint on the borrowing.}, but to apply more realistic conditions, constraints are usually imposed on the cushion to prohibit short-selling of the risky asset (see e.g. \cite{constantinou2009does}). Under these conditions, the cushion
value at any time $t\in[0,T]$ is given by
\begin{equation}\label{maxcushion}
    \displaystyle{C_t = ({V_t}^{\rm CPPI} - F_t)^+,}
\end{equation}
and the total amount invested in the risky asset (which is called the exposure) is obtained by multiplying a constant coefficient, $m$, in the cushion
value
\begin{equation}\label{exposure}
    \displaystyle{E_t = m C_t.}
\end{equation}
It is obvious that higher multiples will result in more profits from stock price increases. Nevertheless, this will also cause faster convergence of the portfolio value to the floor in the case of a dramatic decrease in the stock prices. Note also that different values of the parameter, $m$, change the behavior of the payoff function and $m>1$ provides a convex payoff structure.
From (\ref{maxcushion}) and (\ref{exposure}), it is deduced that if ${V_t}^{\rm CPPI} \leqslant F_t$, then the exposure
will be zero and so the entire portfolio value will be invested in the risk-free asset (see e.g. \cite{cont2009constant} for more details). It could also easily be seen that the value of the portfolio at any time is equal to the current floor plus the cushion value. As cushion is non-negative, the value of the CPPI portfolio is always above the current floor.
\subsection{Constrained CPPI}
In realistic situations, we usually impose a constraint on the portfolio by limiting the exposure to the bounds
$$0<E_t<p{V_t}^{\rm CPPI},$$
in which $p>0$ is a given constant. So the value of the exposure in the constrained CPPI strategy will be obtained as
\begin{equation}\label{EtC}
    \displaystyle{E_t = \min\{mC_t , p{V_t}^{\rm CPPI}\}.}
\end{equation}
Note that in the constrained CPPI case, the portfolio composition is path-dependent (see \cite{boulier2005expected}) and the fraction of the portfolio invested in the risk-free asset is given by
\begin{equation}\label{Bt}
\displaystyle{B_t ={V_t}^{\rm CPPI} - E_t,}
\end{equation}
and the evolution of the portfolio value is described by the following stochastic differential equation (SDE) (see e.g. \cite{boulier2005expected})
\begin{equation}\label{dvt}
\displaystyle{d{V_t}^{\rm CPPI} = E_t \frac{dS_t}{S_t} + ({V_t}^{\rm
CPPI} - E_t ) \frac{dB_t}{B_t}}.
\end{equation}

As mentioned above, we assume that the dynamics of the risky part of the portfolio is given by a regime-switching GBM given by (\ref{rs-gbm}) and so substituting it in (\ref{dvt}), we obtain the equation
\begin{equation}\
    \displaystyle{d{V_t}^{\rm CPPI}= E_t\Big((\mu_{\alpha_t}-r)d_t+\sigma_{\alpha_t} dW_t\Big) + r{V_t}^{\rm CPPI} d_t.}
\end{equation}
Based on restrictions on the exposure given by (\ref{maxcushion}) and (\ref{EtC}), the dynamics of the CPPI portfolio
could be re-written as
\begin{equation}\label{dvtc}
    d{V_t}^{\rm CPPI} = \left\{
\begin{array}{ll}
              \displaystyle{r{V_t}^{\rm CPPI} d_t}, & \hbox{$C_t\leqslant0$,} \\
              \displaystyle{m C_t((\mu_{\alpha_t}-r)d_t+\sigma_{\alpha_t} dW_t) + {V_t}^{\rm CPPI} r d_t}, & \hbox{$0<m C_t<p {V_t}^{\rm CPPI}$,} \\
              \displaystyle{p {V_t}^{\rm CPPI}((\mu_{\alpha_t}-r)d_t+\sigma_{\alpha_t} dW_t) + {V_t}^{\rm CPPI} r d_t}, & \hbox{$p{V_t}^{\rm CPPI} \leqslant m C_t$.} \\
\end{array}
\right.
\end{equation}
By applying It\^{o}'s lemma (see e.g. \cite{oksendal2013stochastic}) and discretizing the dynamics (\ref{dvtc}) at a set of discrete nodes $t_n=n\Delta t$, the solution of this SDE could be approximated by a discrete process of the form
\begin{equation}\label{vtc}
   {V^{\rm CPPI}_{t_{n+1}}} = {V^{\rm CPPI}_{t_{n}}} +  \left\{
\begin{array}{ll}
    \displaystyle{{r V^{\rm CPPI}_{t_{n}}} \Delta t} , & \hbox{$C_{t_n}\leqslant0$,} \\
    \displaystyle{C_{t_n} (m(\mu_{\alpha_{t_n}}-r)+r)\Delta t + m \sigma_{\alpha_{t_n}} \Delta W_{t_n}}, & \hbox{$0<m C_{t_n}<p {V^{\rm CPPI}_{t_n}}$,} \\
    \displaystyle{C_{t_n} (p(\mu_{\alpha_{t_n}}-r)+r)\Delta t + p \sigma_{\alpha_{t_n}} \Delta W_{t_n}}, & \hbox{$p{V^{\rm CPPI}_{t_n}} \leqslant m C_{t_n}$.} \\
\end{array}%
\right.
\end{equation}
\section{VaR-Based Portfolio Insurance}
Similar to the CPPI method, VBPI is a dynamic trading strategy which rebalances the portfolio composition according to
the Value-at-Risk (VaR) concept.
As the value at risk measure concentrates on the downward tail of the return distribution, the VBPI strategy could address the gap risk
by allocating the funds between risky and risk-less assets in such a way as the maximum loss is equated to the VaR at a specified confidence level. This strategy gives a specific discipline for rebalancing portfolios such that the gap risk is controlled.

In the remainder, we assume that the risky part of the portfolio follows a regime-switching geometric Brownian motion as given by (\ref{rs-gbm}).
Let $0\leq w_t \leq 1$ be the fractional allocation of funds to the risk-free asset which will result in $\beta_t=w_t\frac{V_t}{B_t}$ as the number of riskless assets and $\eta_t = (1-w_t)\frac{V_t}{S_t}$ as the number of risky assets in the portfolio. The value of riskless assets grows at a constant risk-free rate according to
\begin{equation}\label{btt}
    \displaystyle{B_t = B_0 \exp(rt),}
\end{equation}
and the value of the risky portfolio will follow a regime-switching dynamics given by
\begin{equation}\label{STrgm}
    \displaystyle{S_t = S_0 \exp\bigg[\sum_{i=1}^H \int_0 ^t\bigg(\mu_i-\frac{\sigma_i^2}{2}\bigg) \delta(i,\alpha_s)ds + \sum_{i=1}^H \int_0^t \sigma_i\delta(i,\alpha_s)dW_s \bigg].}
\end{equation}
In the above expression, $\delta(i,\alpha_s)$ is an indicator function being equal to 1 if we are in the state $\alpha_s=i$ and 0, otherwise. So, the value of the portfolio at time, $t$, is given by
\begin{equation}\label{vt}
    \displaystyle{{V_{t}}^{\rm VBPI} = \beta_t B_{t} + \eta_t S_{t} = w_t V_0 \exp(r{t}) + (1-w_t)V_0 \Big(\frac{S_{t}}{S_0}\Big).}
\end{equation}

According to the dynamic VaR-based approach of Jiang et al. \cite{jiang2009effectiveness}, our goal is to adjust the weights in such a way that the constraint
\begin{equation}\label{var2}
     P({V_t}^{\rm VBPI} \leq F_t) = \alpha,
\end{equation}
will hold in all rebalancing times.
By substituting (\ref{vt}) in (\ref{var2}) and denoting the log-return process by $R_t = \ln {\big(\frac{S_t}{S_0}\big)}$, we could show that (\ref{var2}) is equivalent to
\begin{equation}\label{var3}
    P\bigg( \frac{S_t}{S_0} \leq \frac{F_t-w_tV_0\exp(rt)}{(1-w_t)V_0}\bigg)= P\bigg( R_t \leq
    \ln \bigg( \frac{F_t-w_tV_0\exp(rt)}{(1-w_t)V_0}\bigg)\bigg)= \alpha.
\end{equation}
Equation (\ref{var3}) could be re-written as
\begin{equation}\label{VaRint}
    \int^{\ln \big( \frac{F_t-w_tV_0exp(rt)}{(1-w_t)V_0}\big)}_{-\infty} s f_{R_t}(s) ds = \alpha,
\end{equation}
in which $f_{R_t}(\cdot)$ is the probability density function (pdf) of the log-return process, $R_t$.

According to (\ref{VaRint}), the upper limit in the integral term is the VaR of the random variable $R_t$ at confidence level, $\alpha$, which is defined as the possible maximum loss of a portfolio over a given time horizon within a fixed confidence level and is given by
\begin{equation}\label{var4}
    \ln \Big( \frac{F_t-w_tV_0\exp(rt)}{(1-w_t)V_0}\Big) = q_t.
\end{equation}
So the weight of the risky asset will be obtained as
\begin{equation}\label{var5}
    w_t = \frac{F_t-V_0 \exp(q_t)}{V_0\Big(\exp(rt)- \exp(q_t)\Big)}.
\end{equation}
In the remainder of this section, we describe an efficient way to calculate the value of $q_t$ at each rebalancing time.
\subsection{Details of Calculating Value at Risk}
In order to calculate the VaR in (\ref{var5}) above, we need to know the distribution of $R_t$. As described in Hainaut \cite{hainaut2011risk}, the characteristic function of $R_t$, denoted as $\varphi_t(\cdot)$ and defined by
\begin{equation}\label{chrct1}
    \varphi_t(\vartheta) = \mathbb{E}\big(e^{i\vartheta R_t}\big) =
    \int^{+\infty}_{-\infty} e^{i\vartheta s} f_{R_t} (s)ds,
\end{equation}
could be obtained analytically and so using the characteristic function, we could determine the probability density function of $f_{R_t}$ in a simple and efficient way. Substituting $R_t = \ln\Big(\frac{S_t}{S_0}\Big)$ in (\ref{chrct1}), we could write
\begin{equation}\label{chrct2}
    \varphi_t(\vartheta) = \mathbb{E}\bigg( \Big( {\frac{S_t}{S_0}}\Big) ^{i\vartheta}
    \bigg).
\end{equation}
In order to obtain an analytic expression for $ \varphi_t(\vartheta)$, we need the following result.
\begin{Proposition}[Hainaut  \cite{hainaut2011risk}] Let the matrix, $B_{\gamma}$, be defined by
\begin{equation}\label{Bgama}
 \displaystyle{B_\gamma = Q' + diag \left(%
\begin{array}{c}
  \gamma \bigg(\mu_1 - \frac{\sigma_1^2}{2}\bigg) + \frac{1}{2} \gamma^2 \sigma_1^2 \\
  . \\
  . \\
  . \\
  \gamma \bigg(\mu_H - \frac{\sigma_H^2}{2}\bigg) + \frac{1}{2} \gamma^2 \sigma_H^2 \\
\end{array}%
\right),\quad \forall \gamma \in \mathbb{R}.}
\end{equation}
Then we have
\begin{equation}\label{Ests01}
    \mathbb{E} \bigg(\bigg(\frac{S_t}{S_0}\bigg)^{\gamma}  |
    \mathcal{F}_0 \bigg) = \mathbb{E}(\langle \exp(B_\gamma t)
    \delta(0) ; \textbf{1}\rangle | \mathcal{F}_0) = \sum_{i=1}^M p_i (0)(\langle \exp(B_\gamma
    t) e_i ; \textbf{1}\rangle),
\end{equation}
in which $\delta(t)= (\delta(i,\alpha_t): i\in \mathcal{H})'$ is a vector taking its values in the set of unit vectors $\{e_1,e_2,\cdots,e_H\}$ and $\textbf{1}$ is a vector of $H$ ones.
\end{Proposition}
According to equation (\ref{Ests01}), the characteristic function of $R_t$ could be calculated as
\begin{equation}\label{chrcres}
    \varphi_t(\vartheta)= \sum_{i=1}^H p_i (0)(\langle \exp(B_{i \vartheta} T) e_i ; \textbf{1}\rangle).
\end{equation}
By inverting the Fourier transform, the probability density function of $R_t$ could now be obtained as
\begin{equation}\label{invFF2}
    f_{R_t}(s) = \frac{1}{2 \pi} \int^{+\infty}_{-\infty}\varphi(\vartheta) e^{-i\vartheta
    s}d\vartheta =\frac{1}{\pi} \int^{+\infty}_{0}\varphi(\vartheta) e^{-i\vartheta
    s}d\vartheta.
\end{equation}
As described in Hainaut \cite{hainaut2011risk}, the integral in (\ref{invFF2}) could be calculated by the fast Fourier transform (FFT) method. The output of the FFT algorithm is the distribution function of $R_t$. Using the obtained distribution, the $\alpha$-quantile of $R_t$ which is the desired VaR level could be calculated by
\begin{equation}\label{vtL}
    q_t = w_t V_t \exp(rT) + (1-w_t)V_t \exp(r_\alpha).
\end{equation}
\section{Risk-Adjusted Performance Measures}
Evaluating portfolio performance is a key activity in financial economics \cite{levy2015stochastic,shaked2007stochastic}. During the years, a wide variety of performance measures have been introduced into the field of finance to guide the investors (both private and institutional) in comparing and ranking of investment portfolios and evaluating the added value of portfolio managers (see, e.g. \cite{hendricks1993hot,powell2002capital,grinold2000active,farinelli2008beyond,caporin2014survey}).

There exist a number of performance measures in the literature which focus mainly on the mean and variance of the return distribution
(see e.g. \cite{cogneau2009more} and references therein). A commonly used measures is the Sharpe ratio (see e.g. \cite{sharpe1994sharpe})
calculated as the ratio of the expected excess return of an investment to its return volatility. Originally motivated by
mean-variance analysis and the Capital Asset Pricing Model (CAPM), the Sharpe ratio is routinely used in many different contexts, from
performance attribution to tests of market efficiency and risk management. This measure has been used in various researches in
order to compare different strategies (\cite{annaert2009performance}, \cite{hamidi2009risk}, \cite{hamidi2014dynamic}).

On the other hand, Keating and Shadwick \cite{keating2002universal} and Cascon et al. \cite{cascon2003omega}
have introduced a new performance measure, called ``Omega''  which considers the returns both below and above a given loss threshold,
$L$, selected by the investor.
\subsection{Omega Measure}
By dividing the returns into two classes according to the loss threshold, $L$, the returns below it are
considered as losses and above it as gains. More precisely, let $F_X(x)$ denote the cumulative distribution function (CDF) of the return distribution defined on the interval $(a,b)$. The Omega measure is defined by the expression
\begin{equation}\label{Omega1}
    \displaystyle{\Omega_X (L) = \frac{\int^{b}_{L}(1-F(x))dx}{\int^{L}_{a}F(x)dx}},
\end{equation}
which could equivalently be written in terms of the final portfolio value, $V_T$, as (see e.g. \cite{kazemi2004omega})
\begin{equation}\label{Omega2}
    \displaystyle{\Omega_V (L) = \frac{{\Bbb E}(V_T - L)^+}{{\Bbb E}(L - V_T)^+}}.
\end{equation}
It is obvious that the Omega measure takes account of the entire return distribution while requiring no parametric assumptions on the distribution. It provides an appropriate performance measure to compare different strategies. At any threshold level, $L$, investors prefer the strategies with a higher Omega value \cite{bertrand2011omega}.
\subsection{Kappa Measure}
The other performance measure used to compare the performance of two or more strategies is the ``Kappa'' measure defined by (see e.g. \cite{bertrand2011omega})
\begin{equation}\label{Kappa2}
    \kappa_n (L) = \frac{{\Bbb E}(V_T) - L}{\bigg[{\Bbb E}\big[(L - V_T)^+\big]^n\bigg]^\frac{1}{n}}.
\end{equation}
It is shown that the Sortino ratio (see e.g.
\cite{sortino1994performance}) could be recovered by considering the case $n=2$ in the above definition.

In the context of portfolio insurance, Bertrand and Prigent \cite{bertrand2011omega} have employed both the Omega and Kappa measures to compare the performance of OBPI and CPPI strategies. They show that the CPPI method performs better than the OBPI strategy for jump-diffusion dynamics of the underlying risky asset. We could also mention the work of Zagst and Kraus \cite{zagst2011stochastic} in which they analyze and compare the performance of OBPI and CPPI strategies by means of stochastic dominance criteria. They derive parameter conditions implying the second and third order stochastic dominance of the CPPI strategy.
\subsection{Downside Risk Measures}
A well-studied risk in portfolio insurance, called the ``gap-risk'', is the probability of the portfolio value to fall below the floor and failing to guarantee the desired final amount. This risk is measured by a quantity called the expected shortfall given default (ES) which
describes the amount which is lost if a shortfall occurs (see e.g. \cite{temocin2018constant,balder2009effectiveness}). In order to make it precise, we first define a loss variable, $L_T$, taken to be  the amount the portfolio value is below the guarantee level at maturity and expressed as
\begin{equation}\label{L_T}
    L_T=[F_T - V_T | F_T<V_T].
\end{equation}
Expected value (expected shortfall) of the loss is a measure that estimates the average amount of the loss when the value of the portfolio
at the maturity is below the guarantee amount (see, e.g. \cite{khuman2008constant})
\begin{equation}\label{ESH}
    {\Bbb E}[L_T]={\Bbb E}[G-V_T|V_T<F_T].
\end{equation}
A portfolio insurance strategy incurs a shortfall (breaks through the floor), if $V_t<F_t$ occurs during the investment horizon.
Percentage of times a loss occurs, when measured over a large number of simulations, $M$, could be estimated by
\begin{equation}\label{Pr}
    {\Bbb P}[L_T]\approx\frac{1}{M}\sum_{j=1}^{M}\mathbbm{1}_{\{V_T(\omega_j)<F_T\}},
\end{equation}
and could be interpreted as the probability that the PI strategy to experience a loss.
Note that ${\Bbb P}[L_T]$ could be interpreted from the buyers perspective as the probability that they will receive only the guarantee at maturity.
\section{Numerical Experiments}
In this section, we study both the VBPI and CPPI strategies with daily, weekly and monthly rebalancing frequencies of the portfolio
between a risky and risk-free asset for a planning horizon of $T = 1$ years\footnote{We assume $260$ trading days and $52$ weeks in a typical year.}. We take the portfolio initial level to be $V_0 = 100$ and the guaranteed level at the maturity to be $F_T = 100$ (i.e. $\pi=1$). We also let the yield of the bond to be constant at an annual rate of $r=4\%$. We assume that the risky asset is driven by a time-inhomogeneous Markov-modulated diffusion process with two distinct regimes (considered here as
bullish and bearish markets) with parameter values for each regime as
$$\mu_{1} = 0.14, \quad \sigma_{1} = 0.16 \quad \ \mu_{2}=-0.01 \quad \sigma_{2} = 0.2.$$
The generator matrix of the underlying Markov process is assumed to be given by
$$Q=\left(%
\begin{array}{cc}
  -0.25 & 0.25 \\
  0.25 & -0.25 \\
\end{array}%
\right).$$
In order to demonstrate the effect of regime shifts on the performance of the CPPI and VBPI strategies, we have conducted a Monte-Carlo simulation study by generating sample paths from the underlying asset's risky dynamics.


To make the two strategies comparable in different scenarios, multiples in the CPPI strategy are chosen such that the strategy's initial
allocation to the risky asset is the same as that under the corresponding VBPI strategy (see e.g. \cite{jiang2009effectiveness}). Using
(\ref{var5}) and the relation
\begin{equation}\label{mequ}
  mC_0 = (1-w_0)V_0,
\end{equation}
the value of $m$ could be calculated.
In the above equation, the weight of the risky asset in the CPPI strategy is assumed the same as the weight of risky asset in the VBPI strategy at the initial time.



In Figures 1-3, we have demonstrated the Omega performance measure as a function of different threshold levels. As pointed out by Bacmann and Scholz \cite{bacmann2003alternative}, Omega involves all the moments of the return distribution including skewness and kurtosis and so, it is an appropriate indicator of the effectiveness of insurance strategies (see also \cite{jiang2009effectiveness}). In all figures and for high threshold levels, the VBPI strategy has a better performance than the CPPI method while in low thresholds, CPPI has a better Omega measure. Also by increasing the confidence level (CL), the Omega measure increases, showing that the performance of the portfolios in higher confidence levels are improved.

The histogram (frequency distribution) of the terminal values of CPPI and VBPI portfolios for daily, weakly and monthly rebalancing periods is depicted in Figures 4-6. They show that the left tail of the frequency distribution for the CPPI strategy is shorter than VBPI. Also by increasing the confidence level, the left tail becomes shorter in both cases. They collectively show that the performance of CPPI strategy is better than VBPI.

In Figure 7, we have plotted the expected final portfolio levels under the VBPI and CPPI strategies (denoted respectively by EVVBPI and EVCPPI) versus the rebalancing period and confidence level. As it is evident from these figures, the expected value increases in both cases as we increase the rebalancing period and decreases as we increase the confidence level. Figure 8 which is concerned with the standard deviation of the final portfolio levels versus the rebalancing period and confidence level, shows a similar behavior but here the CPPI method is indifferent to changing the rebalancing period. In Figure 9, we have plotted the normalized number of portfolios which fail to give the floor value (shortfall probability) versus the rebalancing period and confidence level. In this case, increasing the confidence level will result in decreasing the number of portfolios which fail to give the floor value. Here again we conclude that the CPPI performance is better than VBPI for low confidence levels.

In Figure 10, we have compared the expected terminal value of the two strategies. It shows that the initial behavior of the two strategies are the same, however by increasing the rebalancing period, the value of VBPI portfolio attains higher levels.

%

In Table 1 we have compared the two strategies in terms of Omega and Kappa performance measures for different threshold levels. We observe that for all threshold values, daily rebalancing leads to better results. Among different confidence levels and for daily rebalancing, the 99 percent confidence level will lead to better results while at weekly and monthly rebalancing periods, lower confidence levels will lead to better results. Table 2 also compares the two strategies in terms of the Sharpe ratio. For daily rebalancing, we obtain better results in comparison to the weekly and monthly rebalancing periods.

It is evident from the above experiments that by increasing the confidence level, the return and the standard deviation of the corresponding
portfolios decreases, as expected. Also by increasing the rebalancing period, the performance of both strategies degrades. In all the presented results, the CPPI strategy has a better performance than the VBPI method except for the Omega measure in which for high threshold levels, the VBPI strategy has a slightly better performance.


\begin{figure}
     \centering
    \begin{subfigure}[t]{0.32\textwidth}
        \raisebox{-\height}{\includegraphics[width=\textwidth]{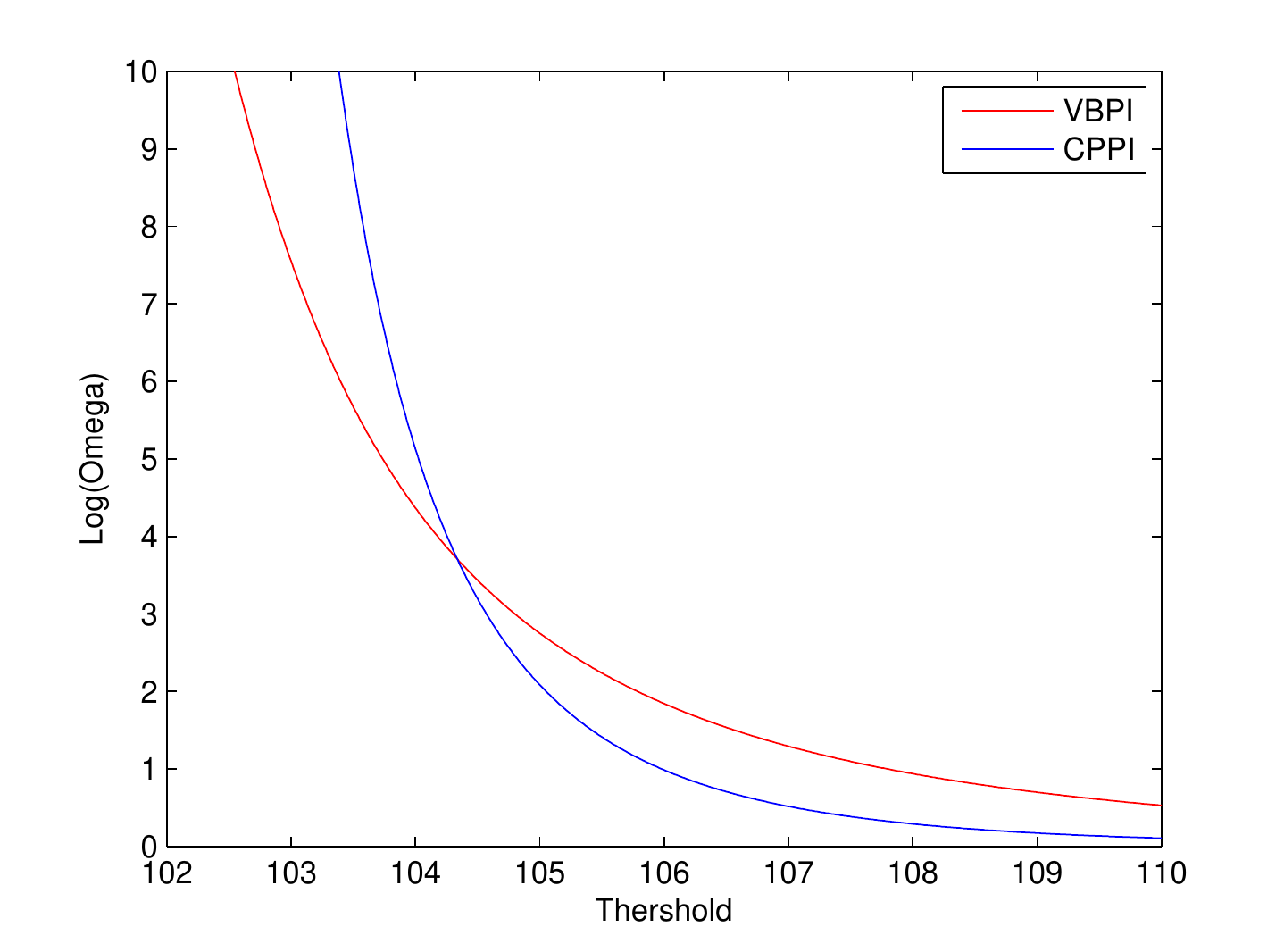}}
        \caption{Daily}
    \end{subfigure}
    \hfill
    \begin{subfigure}[t]{0.32\textwidth}
        \raisebox{-\height}{\includegraphics[width=\textwidth]{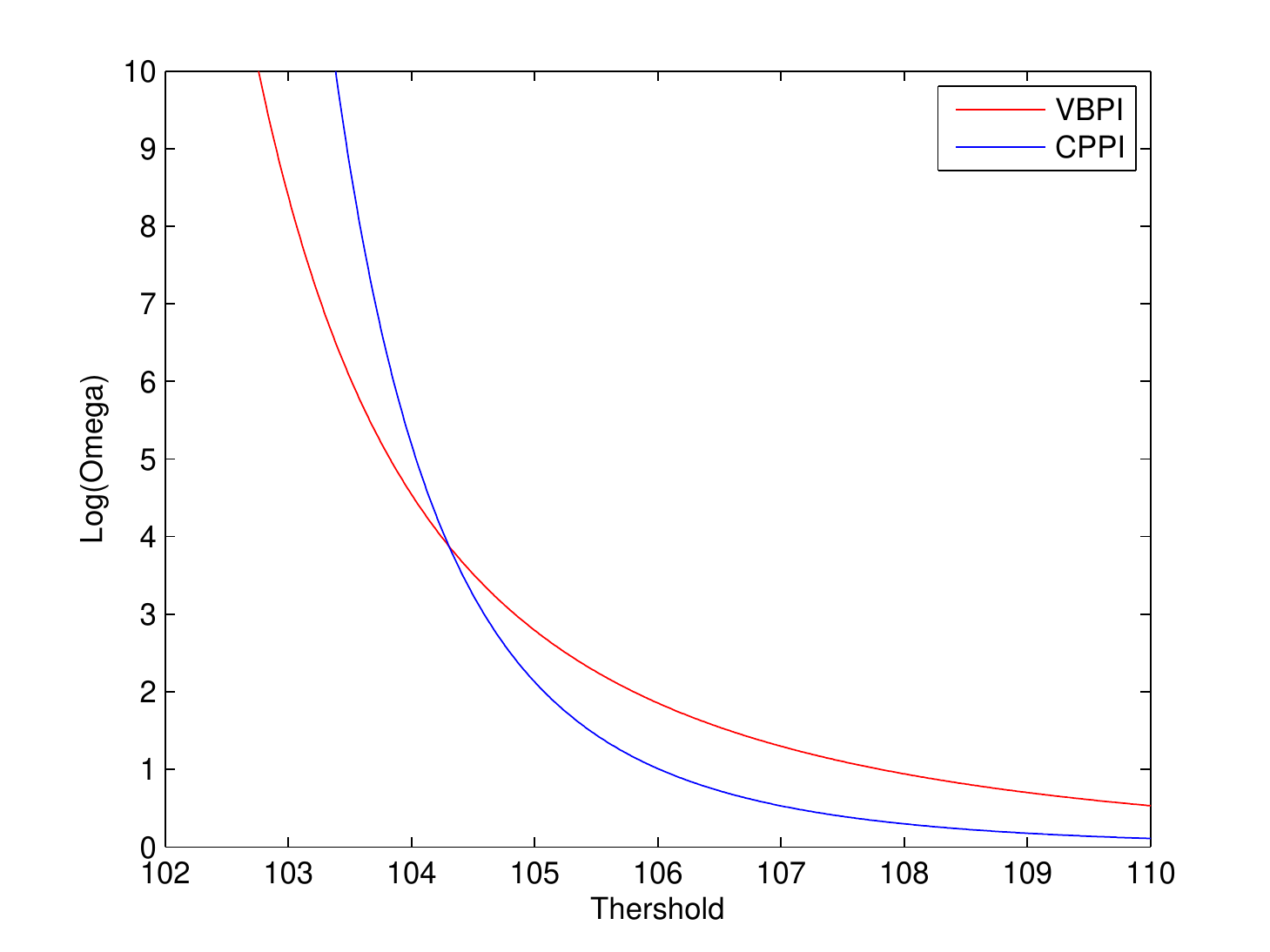}}
        \caption{Weekly}
    \end{subfigure}
    \hfill
    \begin{subfigure}[t]{0.32\textwidth}
        \raisebox{-\height}{\includegraphics[width=\textwidth]{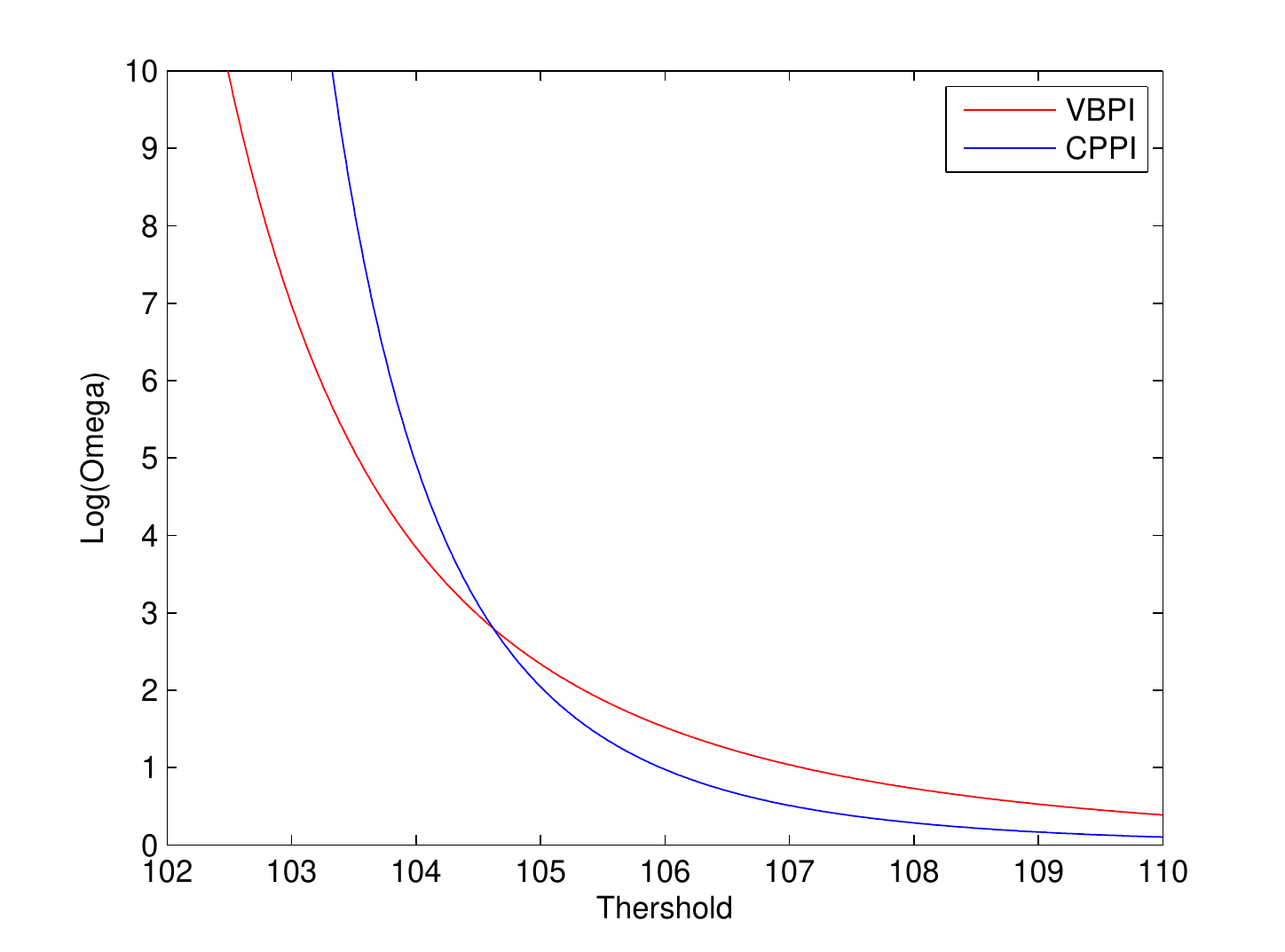}}
        \caption{Monthly}
    \end{subfigure}
    \caption{Omega measure of CPPI and VBPI portfolio at $90\%$ CL.}
\end{figure}
\begin{figure}
     \centering
    \begin{subfigure}[t]{0.32\textwidth}
        \raisebox{-\height}{\includegraphics[width=\textwidth]{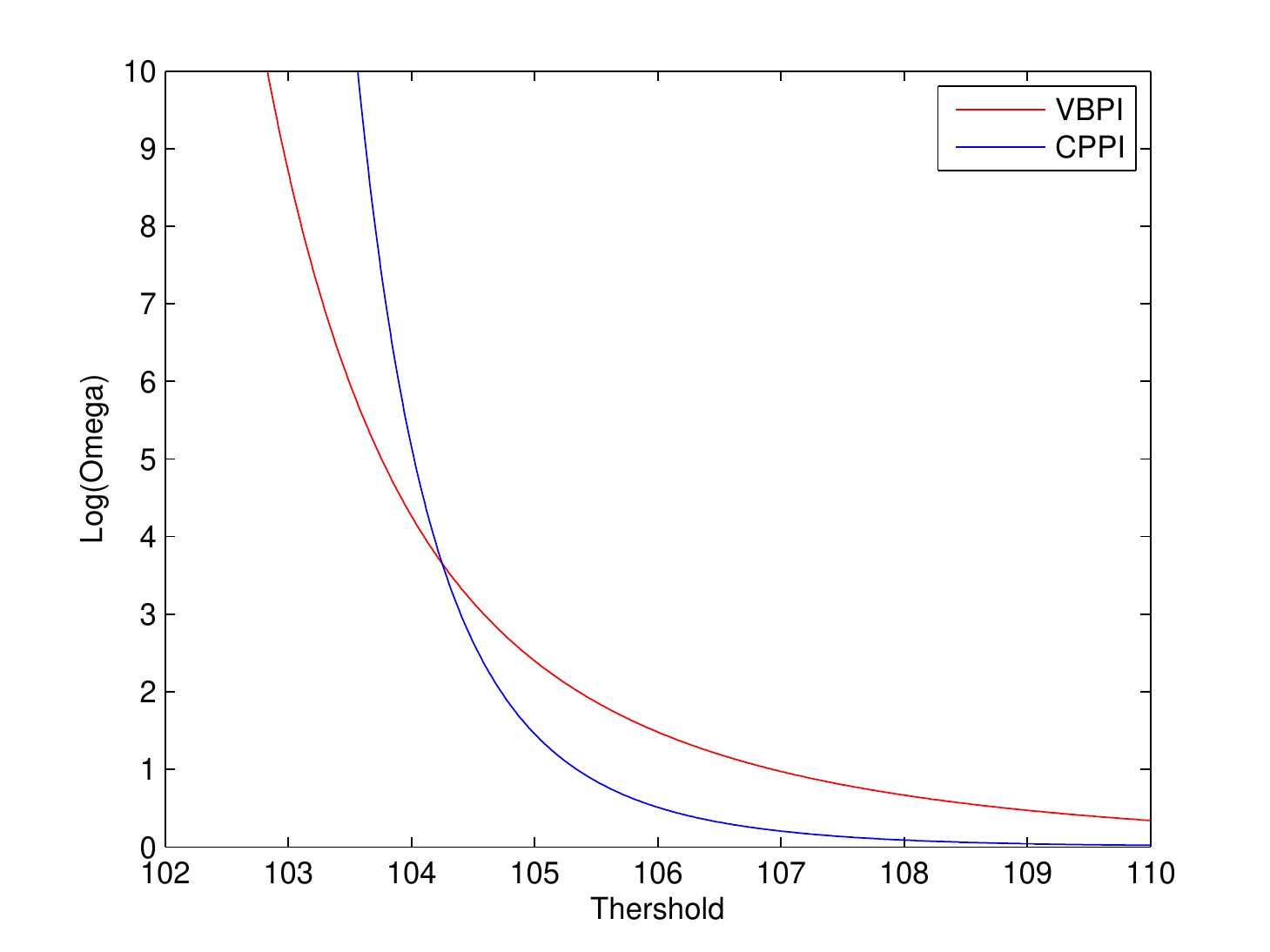}}
        \caption{Daily}
    \end{subfigure}
    \hfill
    \begin{subfigure}[t]{0.32\textwidth}
        \raisebox{-\height}{\includegraphics[width=\textwidth]{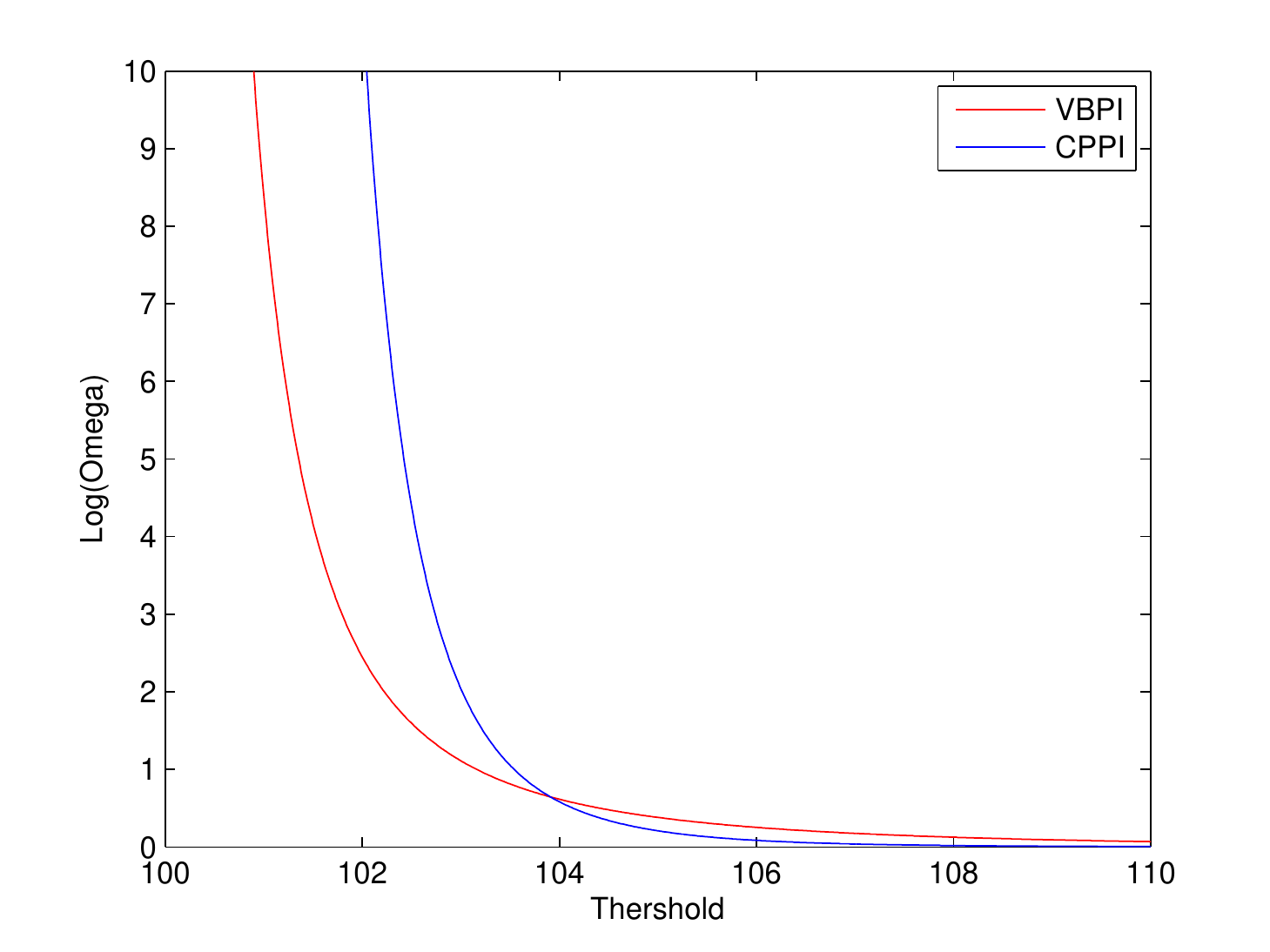}}
        \caption{Weekly}
    \end{subfigure}
    \hfill
    \begin{subfigure}[t]{0.32\textwidth}
        \raisebox{-\height}{\includegraphics[width=\textwidth]{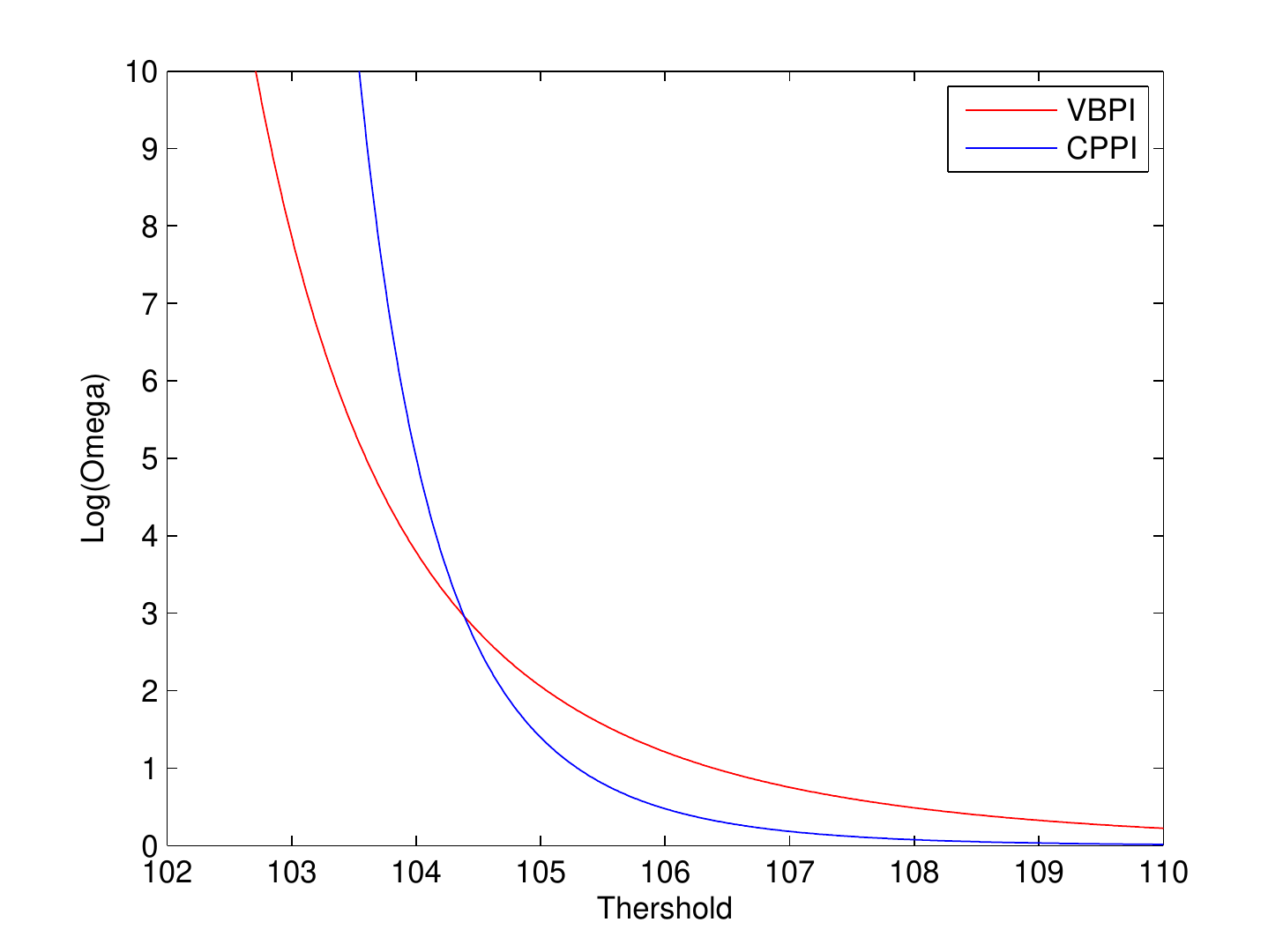}}
        \caption{Monthly}
    \end{subfigure}
    \caption{Omega measure of CPPI and VBPI portfolio at $95\%$ CL.}
\end{figure}
\begin{figure}
     \centering
    \begin{subfigure}[t]{0.32\textwidth}

        \raisebox{-\height}{\includegraphics[width=\textwidth]{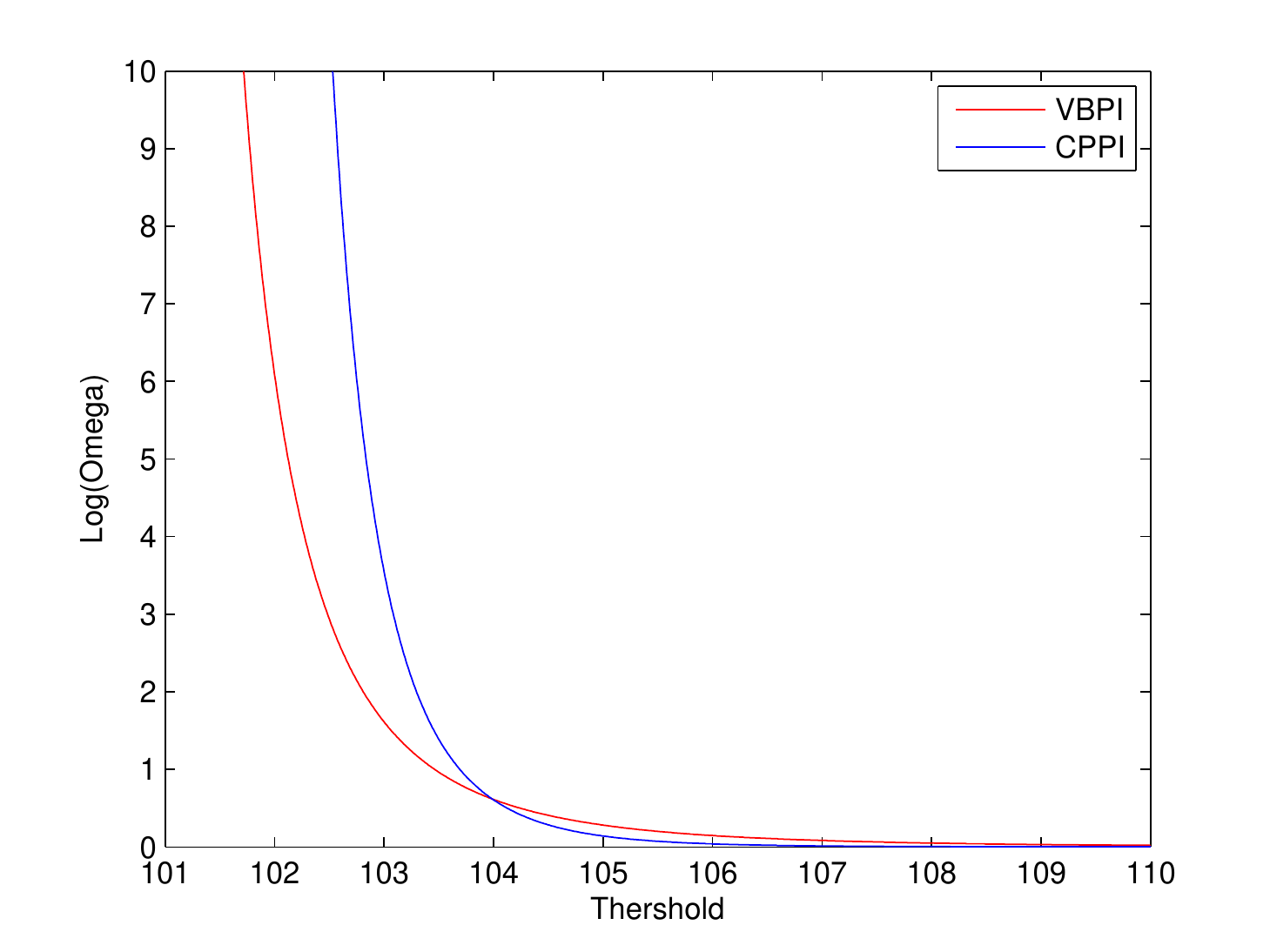}}
        \caption{Daily}
    \end{subfigure}
    \hfill
    \begin{subfigure}[t]{0.32\textwidth}
        \raisebox{-\height}{\includegraphics[width=\textwidth]{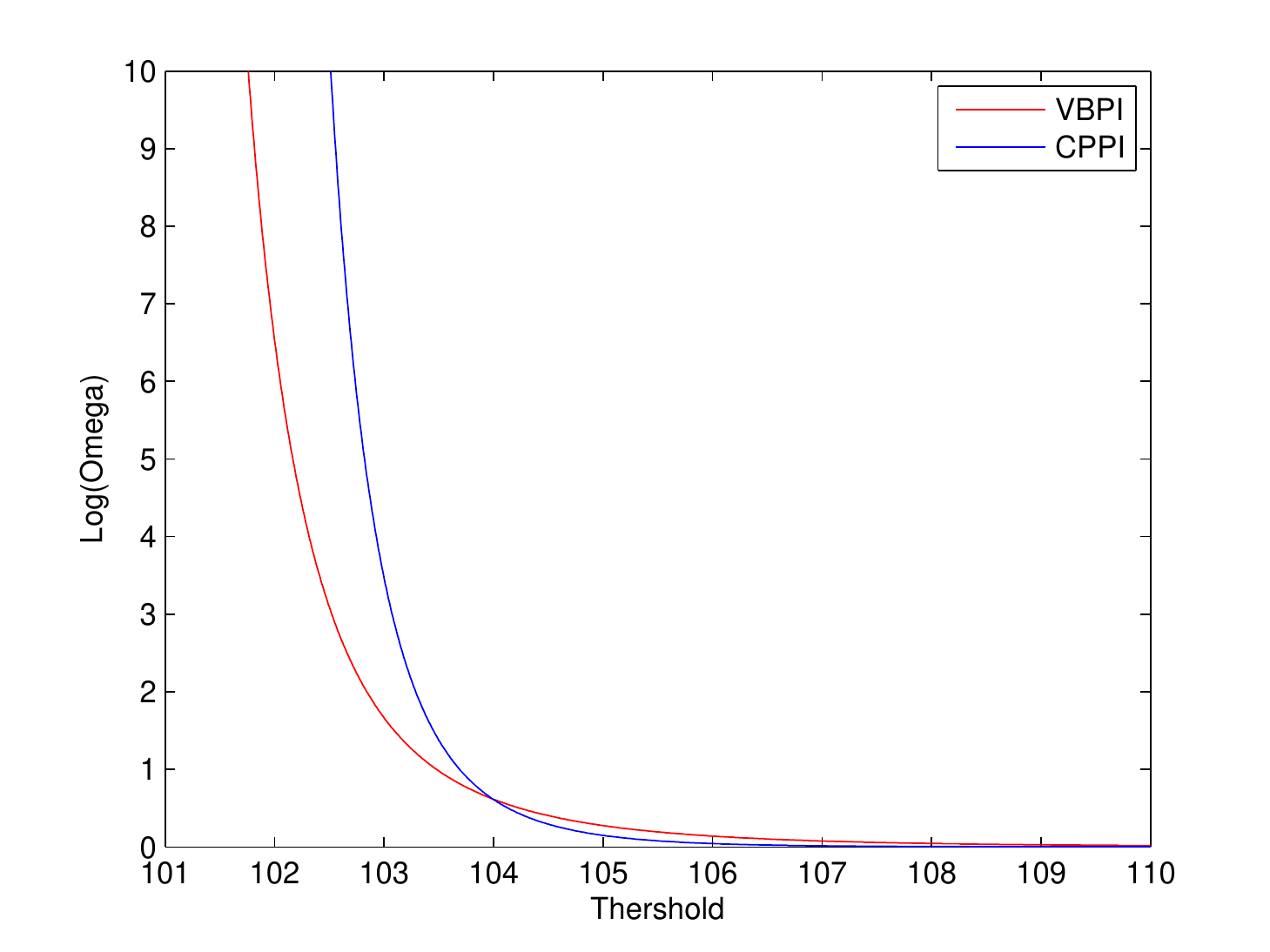}}
        \caption{Weekly}
    \end{subfigure}
    \hfill
    \begin{subfigure}[t]{0.32\textwidth}
        \raisebox{-\height}{\includegraphics[width=\textwidth]{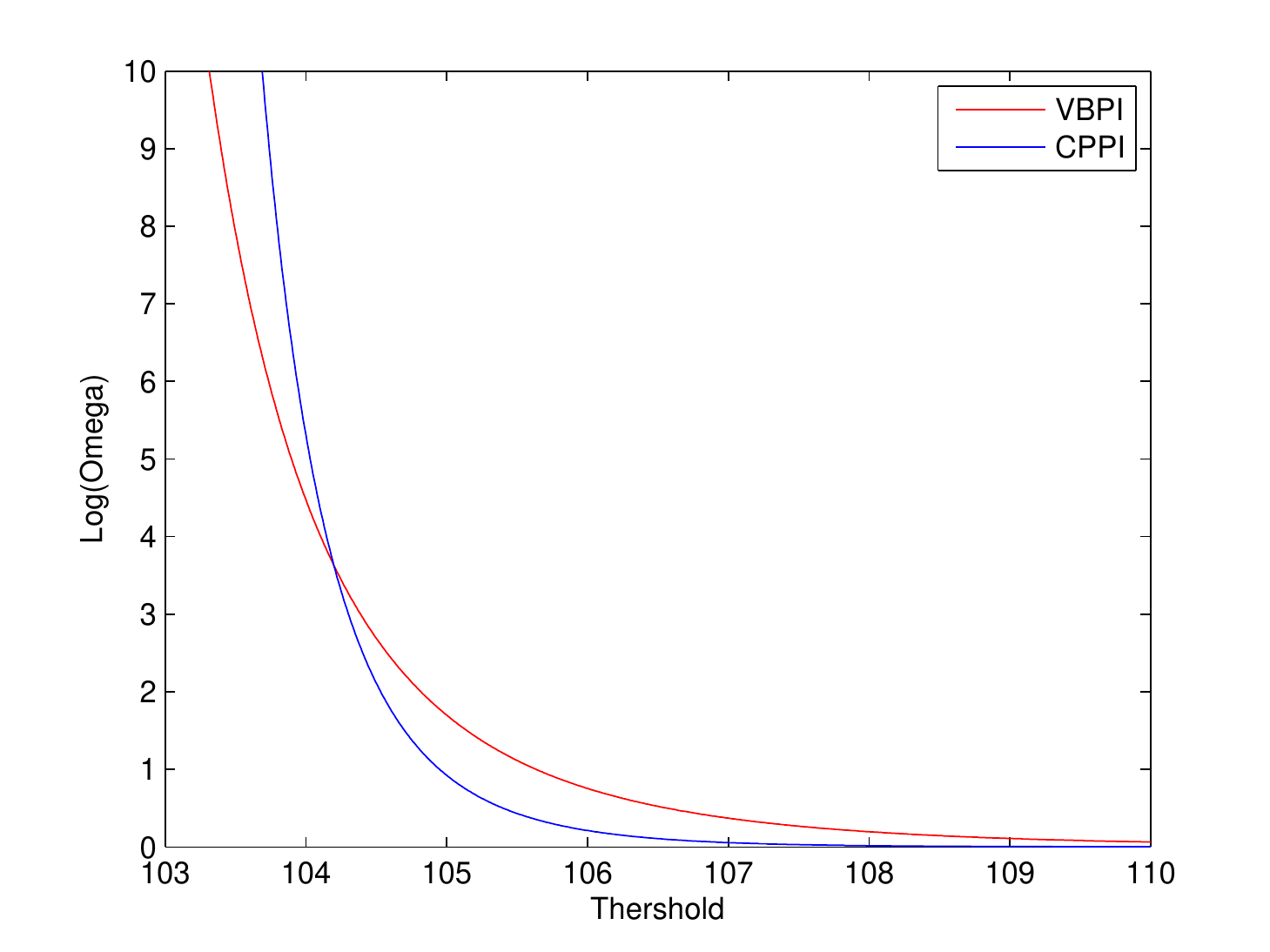}}
        \caption{Monthly}
    \end{subfigure}
    \caption{Omega measure of CPPI and VBPI portfolio at $99\%$ CL.}
\end{figure}
\begin{figure}
     \centering
    \begin{subfigure}[t]{0.32\textwidth}
        \raisebox{-\height}{\includegraphics[width=\textwidth]{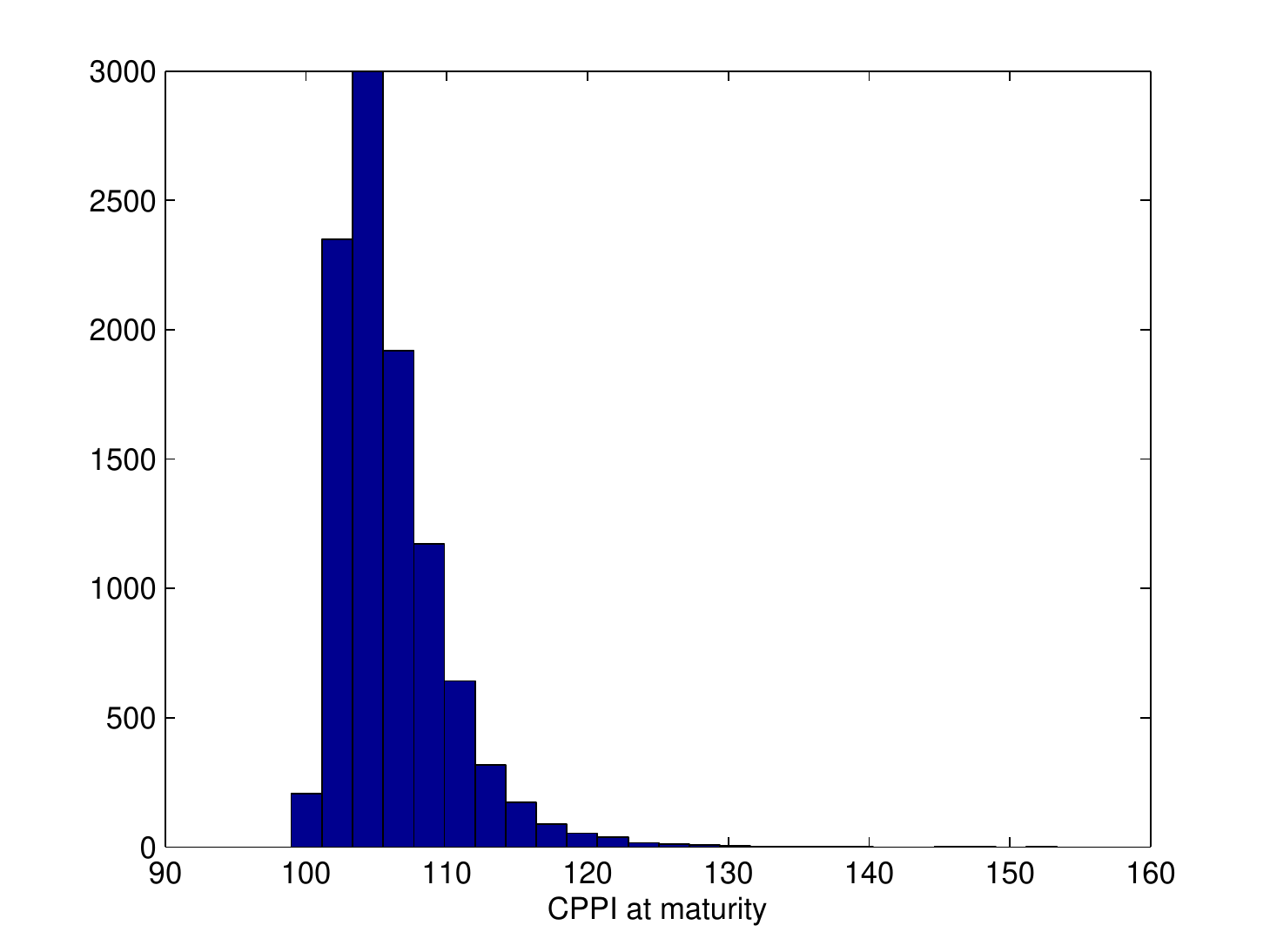}}
        \caption{CPPI-Daily-CL 90\% }
        \raisebox{-\height}{\includegraphics[width=\textwidth]{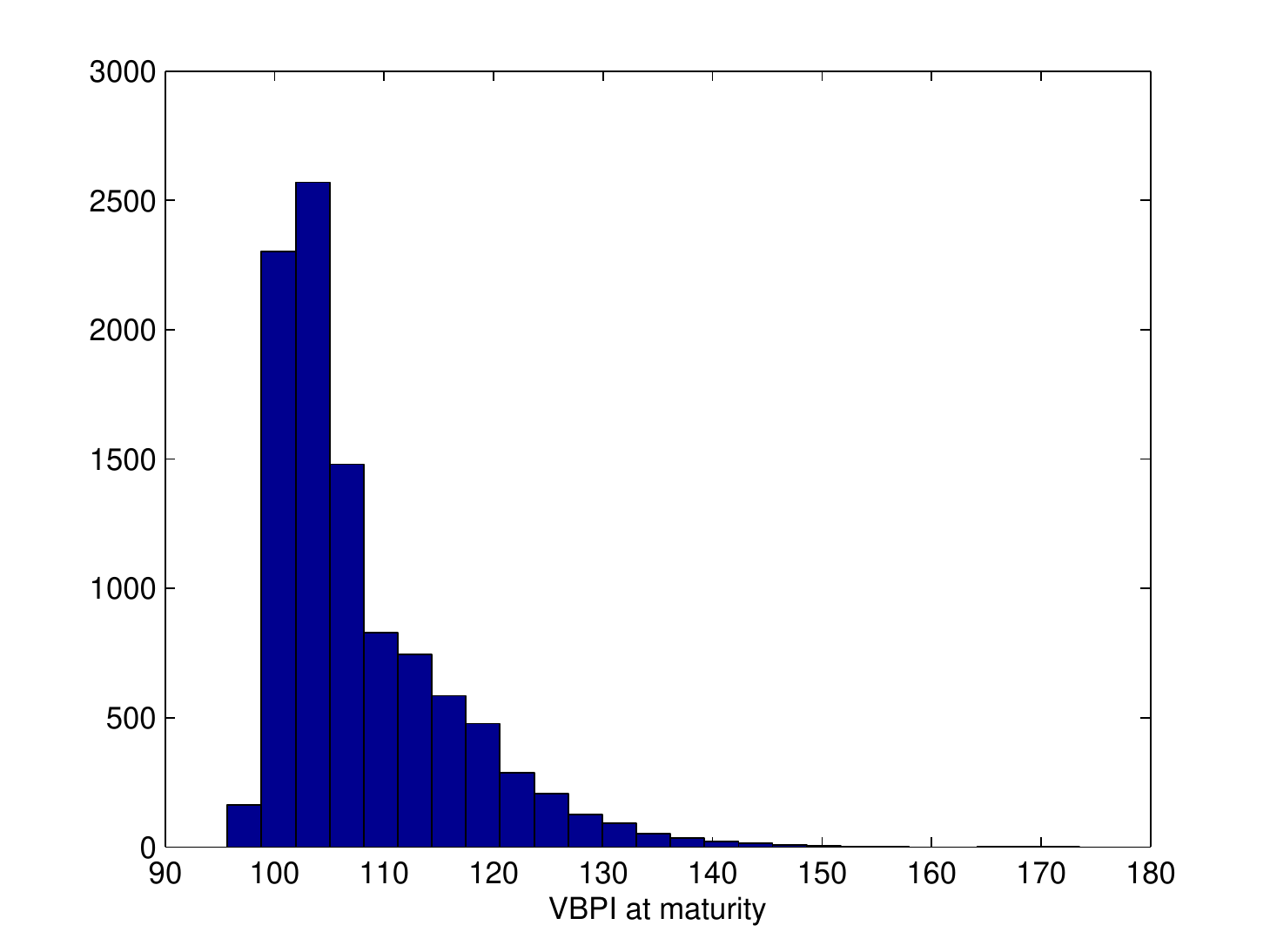}}
        \caption{VBPI-Daily-CL 90\% }
    \end{subfigure}
    \hfill
    \begin{subfigure}[t]{0.32\textwidth}
        \raisebox{-\height}{\includegraphics[width=\textwidth]{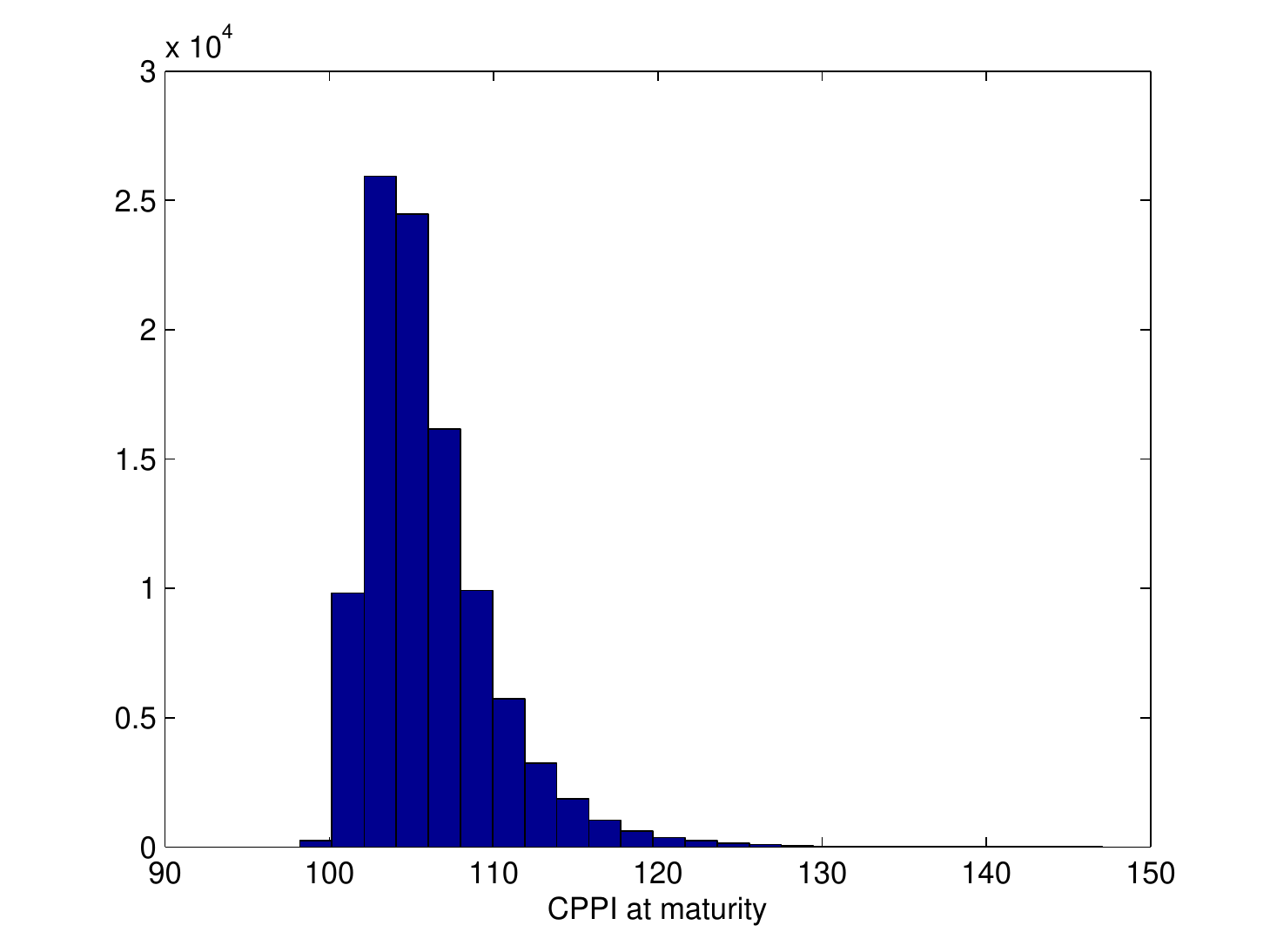}}
        \caption{CPPI-Weekly-CL 90\% }
        \raisebox{-\height}{\includegraphics[width=\textwidth]{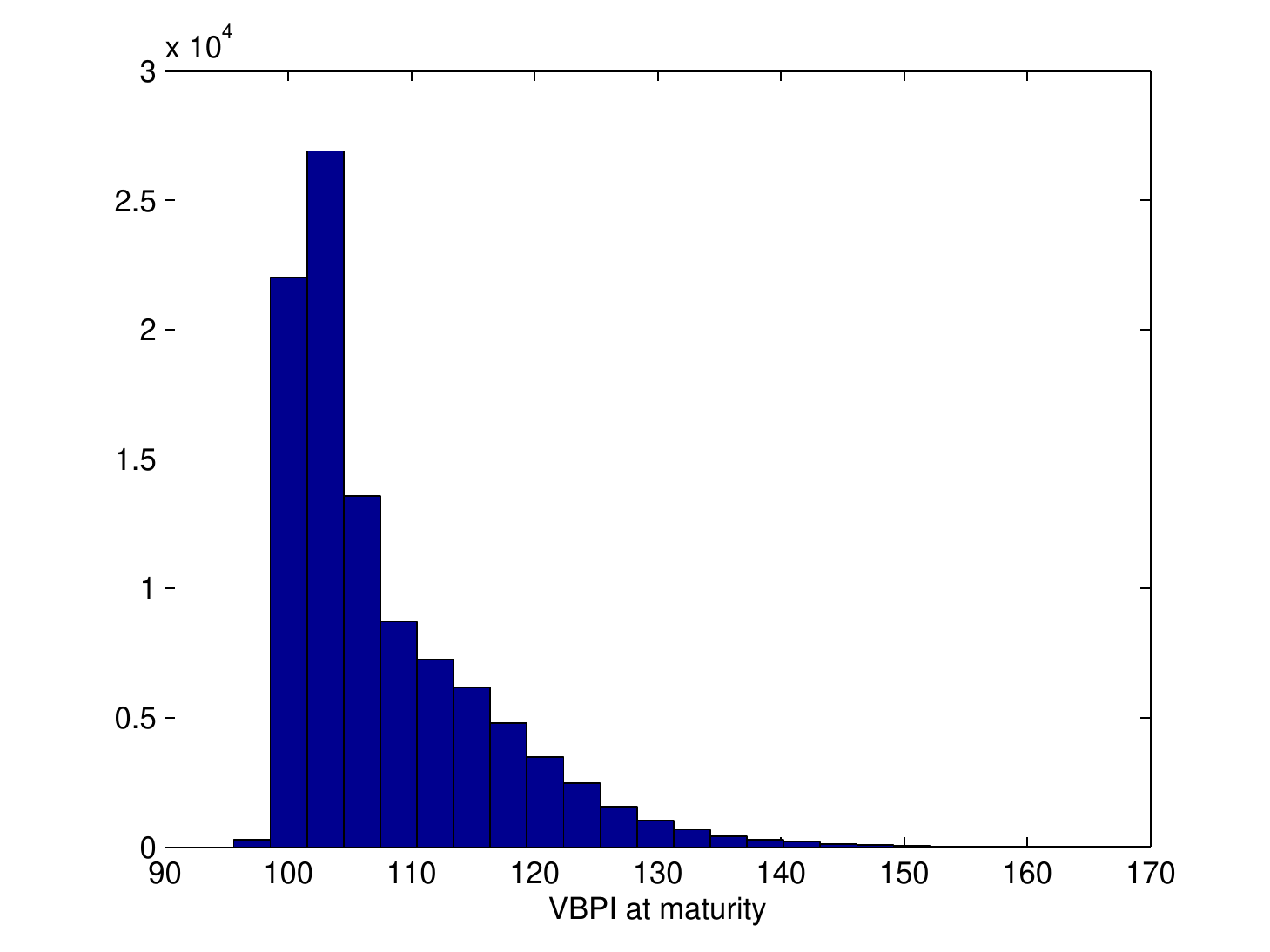}}
        \caption{VBPI-Weekly-CL 90\% }
    \end{subfigure}
    \hfill
    \begin{subfigure}[t]{0.32\textwidth}
        \raisebox{-\height}{\includegraphics[width=\textwidth]{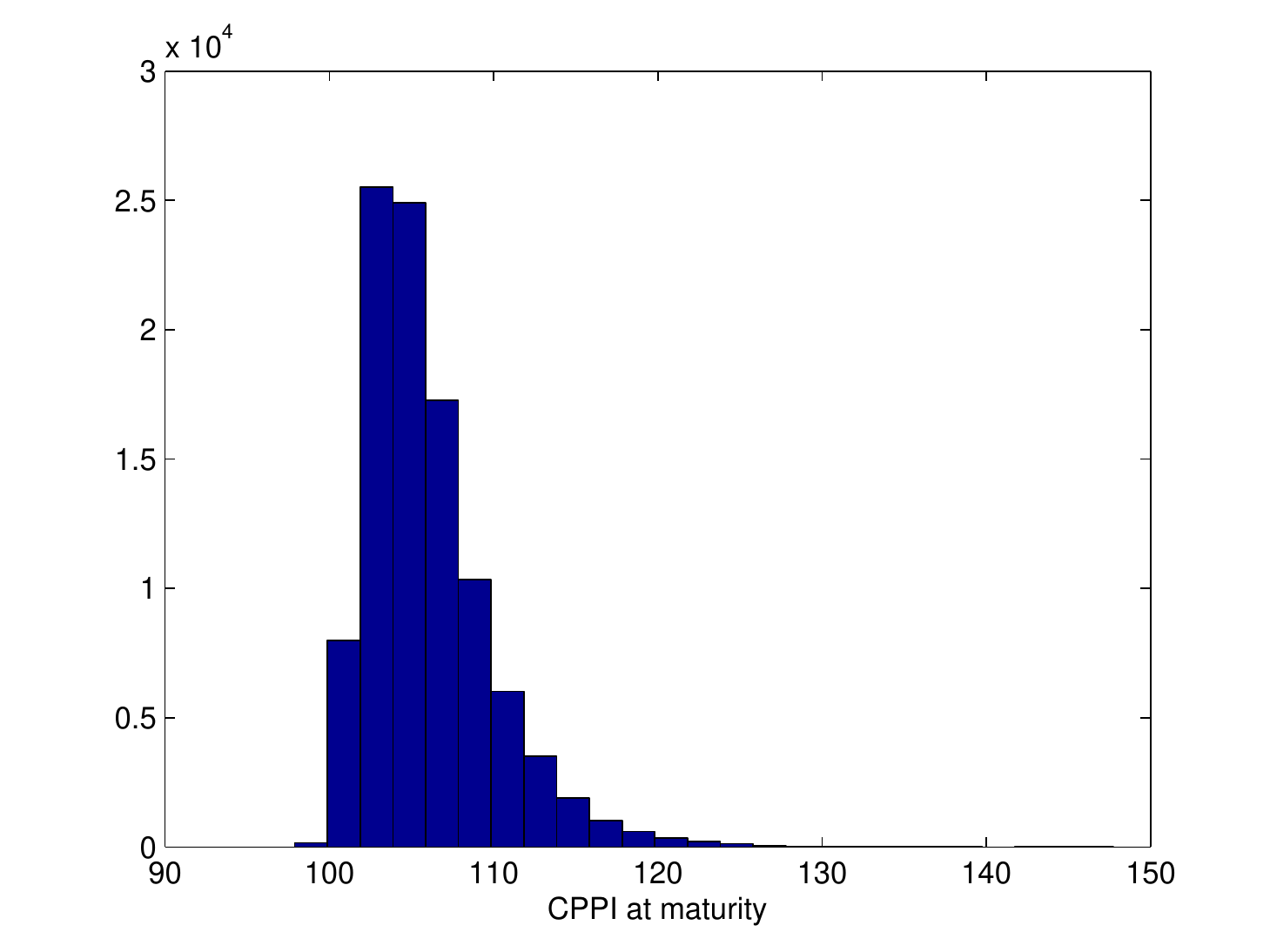}}
        \caption{CPPI-Monthly-CL 90\% }
        \raisebox{-\height}{\includegraphics[width=\textwidth]{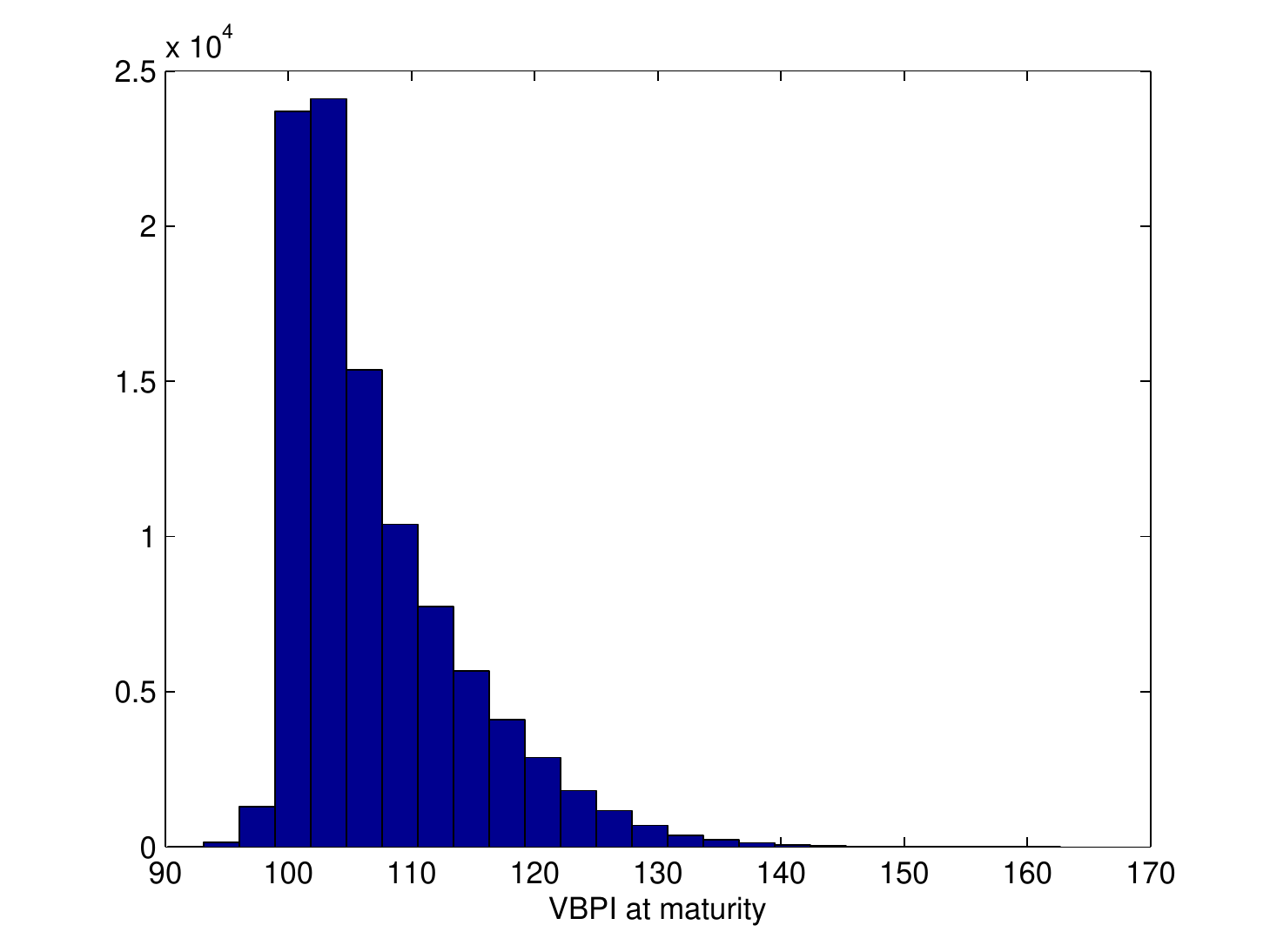}}
        \caption{VBPI-Monthly-CL 90\% }
    \end{subfigure}
        \caption{Frequency of the terminal value of the CPPI and VBPI portfolios at $90\%$ CL.}
\end{figure}

\begin{figure}
     \centering
    \begin{subfigure}[t]{0.32\textwidth}
        \raisebox{-\height}{\includegraphics[width=\textwidth]{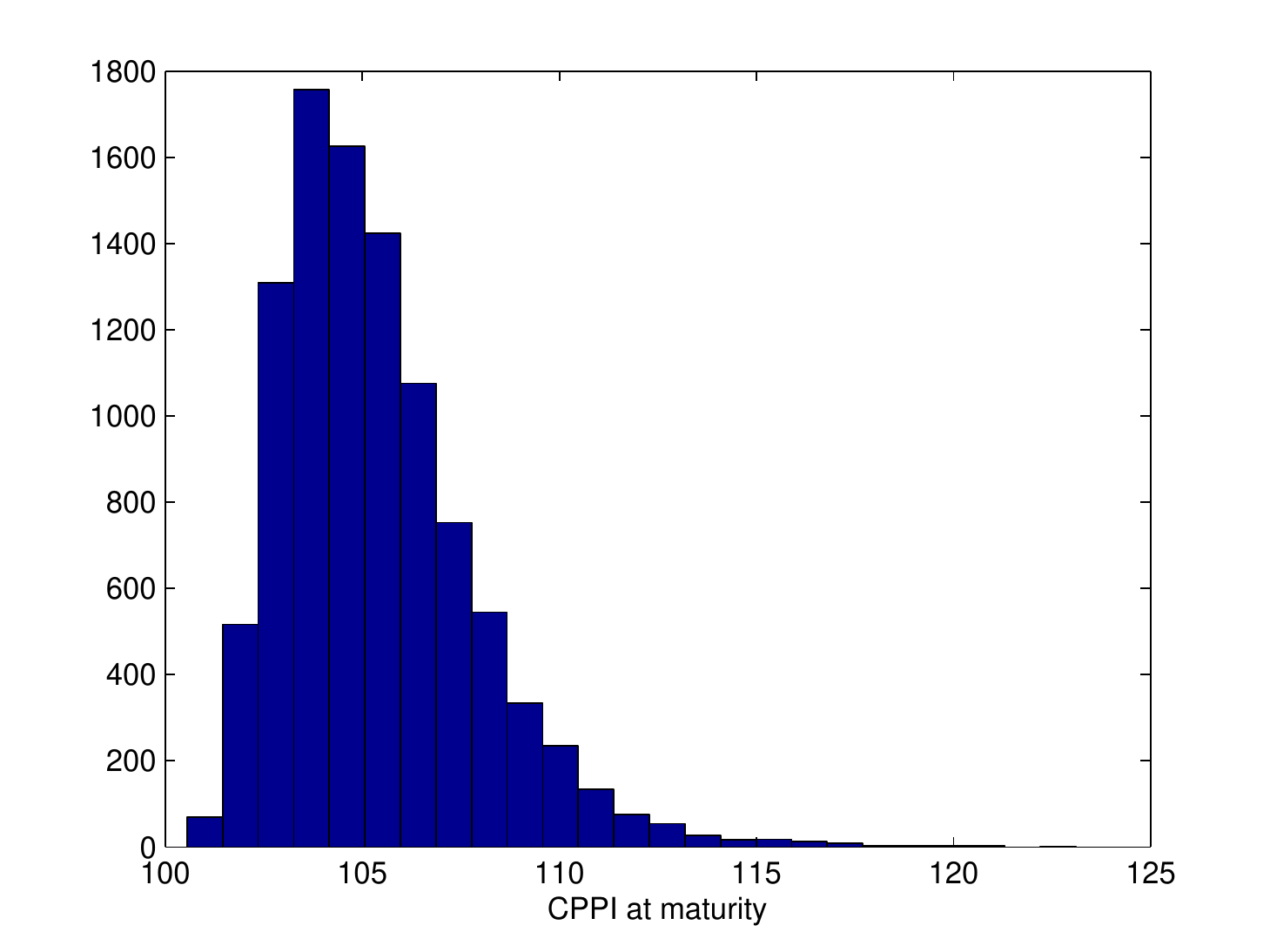}}
        \caption{CPPI-Daily-CL 95\% }
        \raisebox{-\height}{\includegraphics[width=\textwidth]{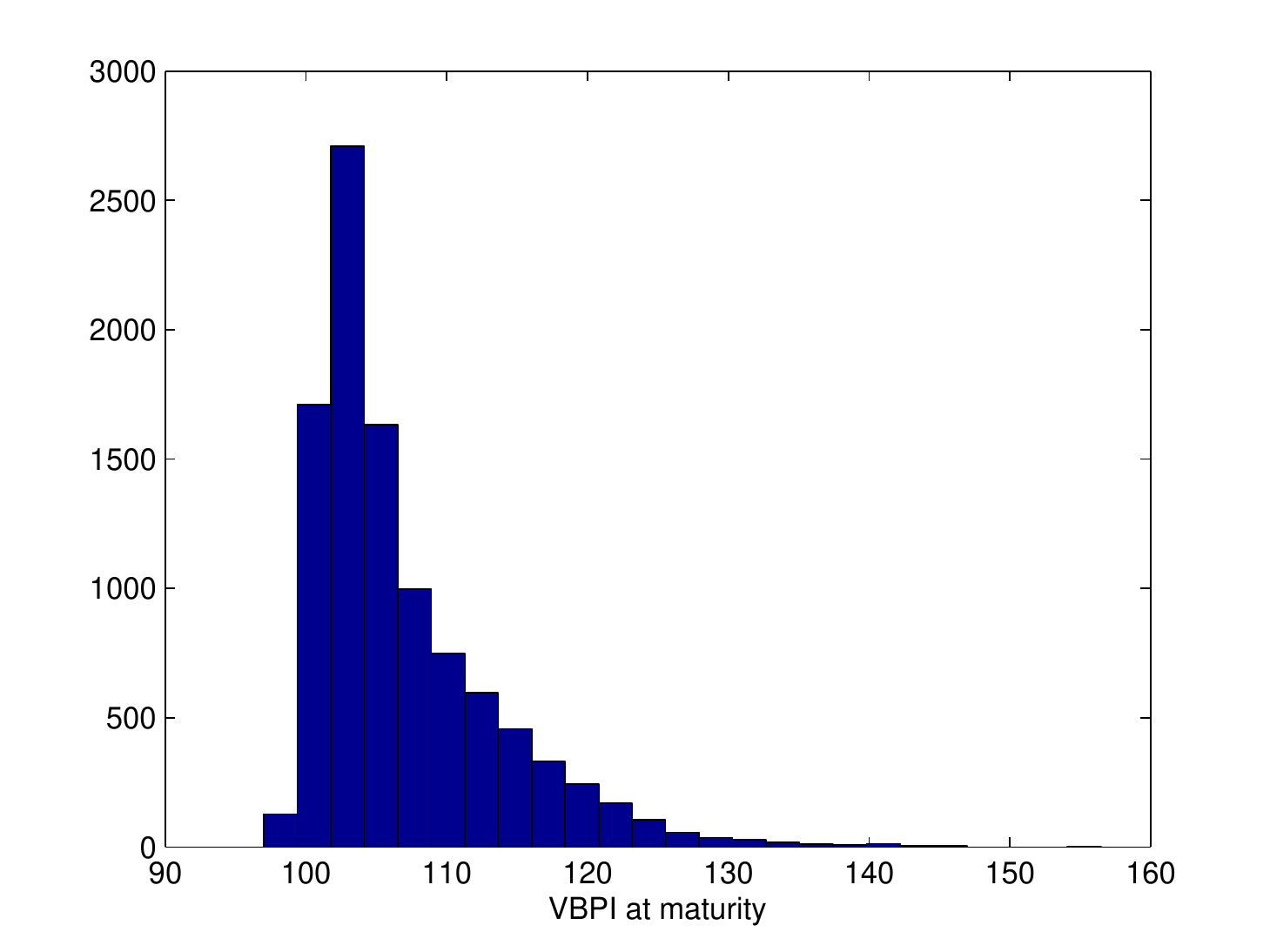}}
        \caption{VBPI-Daily-CL 95\% }
    \end{subfigure}
    \hfill
    \begin{subfigure}[t]{0.32\textwidth}
        \raisebox{-\height}{\includegraphics[width=\textwidth]{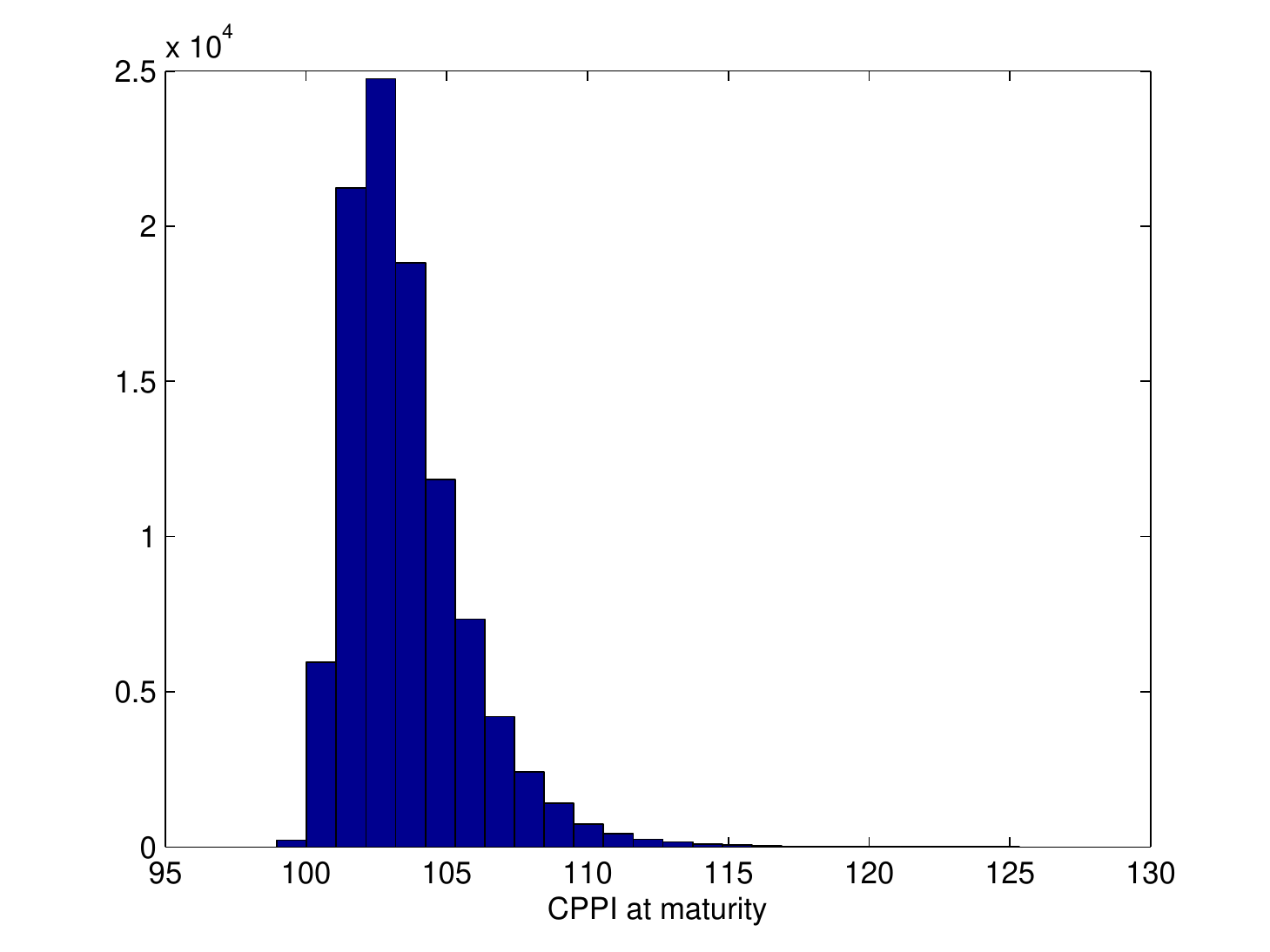}}
        \caption{CPPI-Weekly-CL 95\% }
        \raisebox{-\height}{\includegraphics[width=\textwidth]{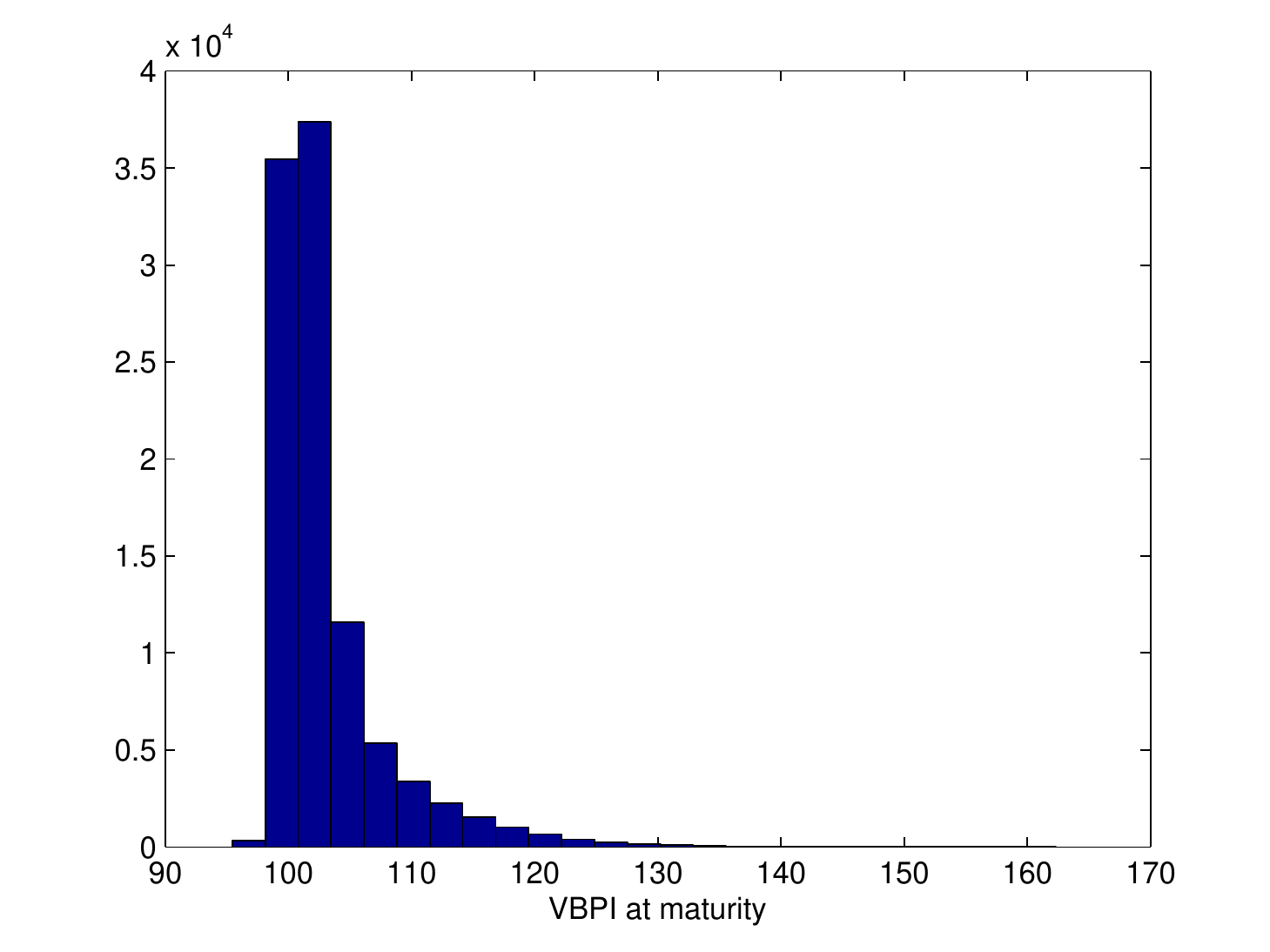}}
        \caption{VBPI-Weekly-CL 95\% }
    \end{subfigure}
    \hfill
    \begin{subfigure}[t]{0.32\textwidth}
        \raisebox{-\height}{\includegraphics[width=\textwidth]{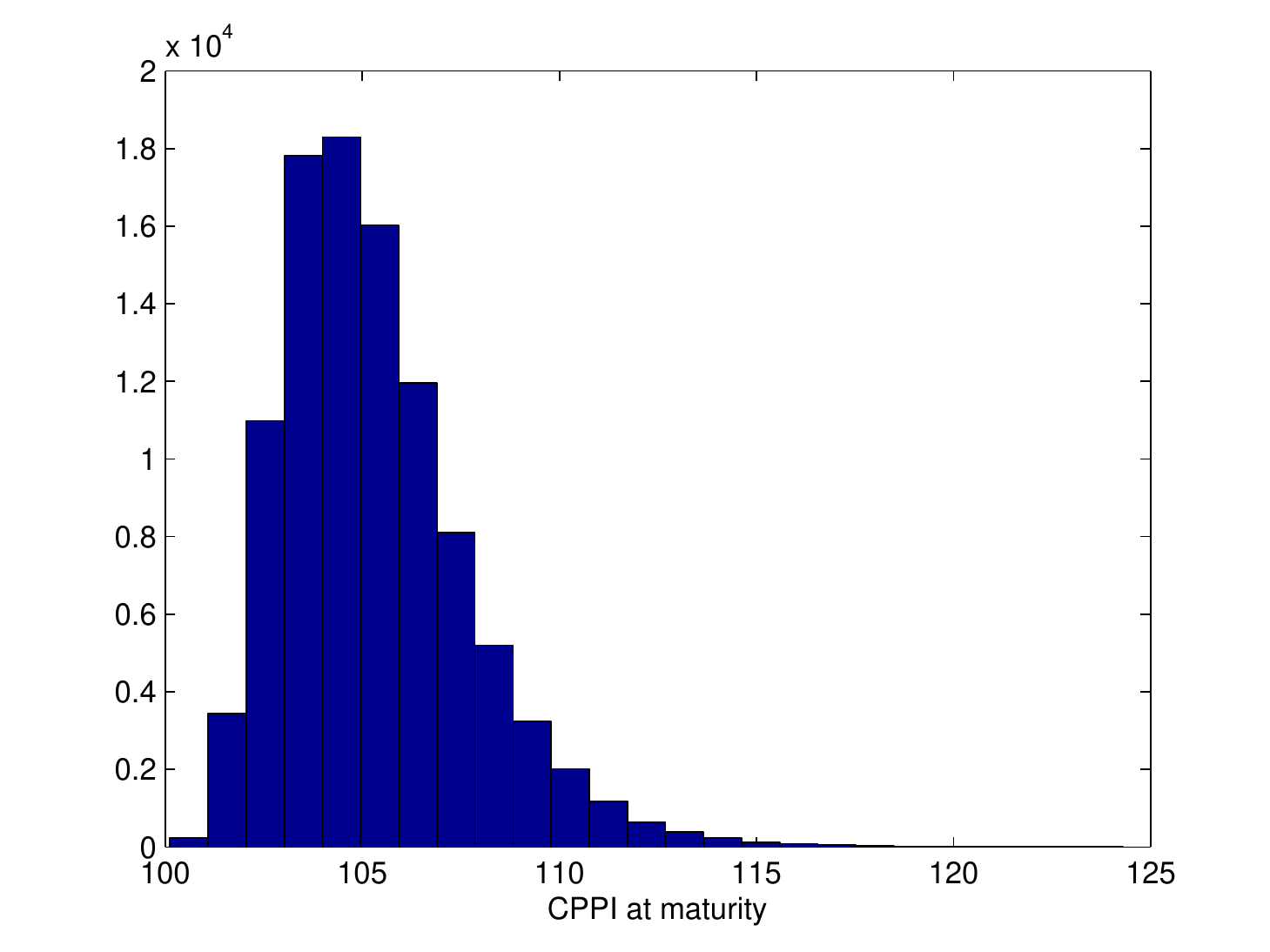}}
        \caption{CPPI-Monthly-CL 95\% }
        \raisebox{-\height}{\includegraphics[width=\textwidth]{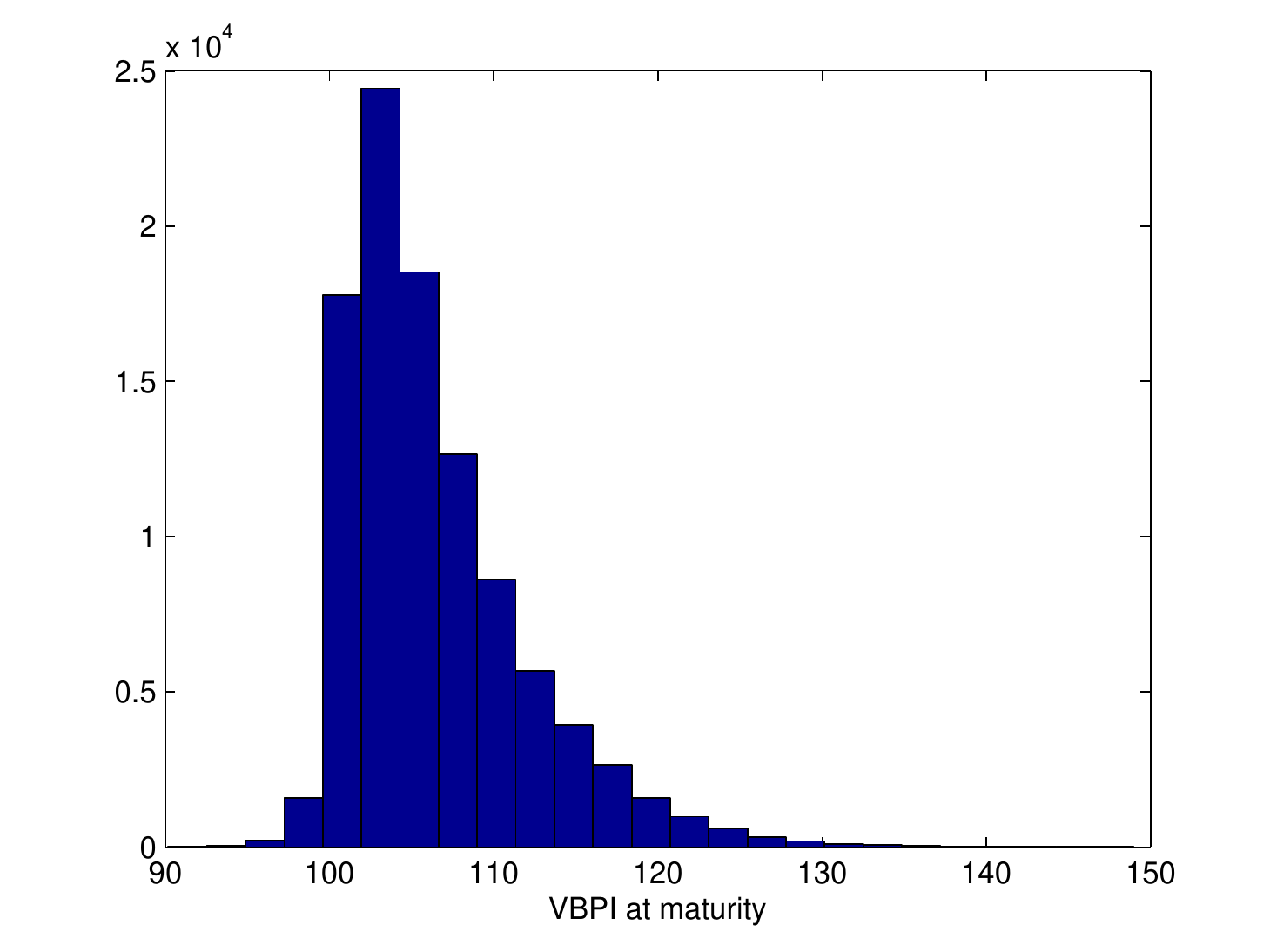}}
        \caption{VBPI-Monthly-CL 95\% }
    \end{subfigure}
    \caption{Frequency of the terminal value of the CPPI and VBPI portfolios at $95\%$ CL}
\end{figure}

\begin{figure}
     \centering
    \begin{subfigure}[t]{0.32\textwidth}
        \raisebox{-\height}{\includegraphics[width=\textwidth]{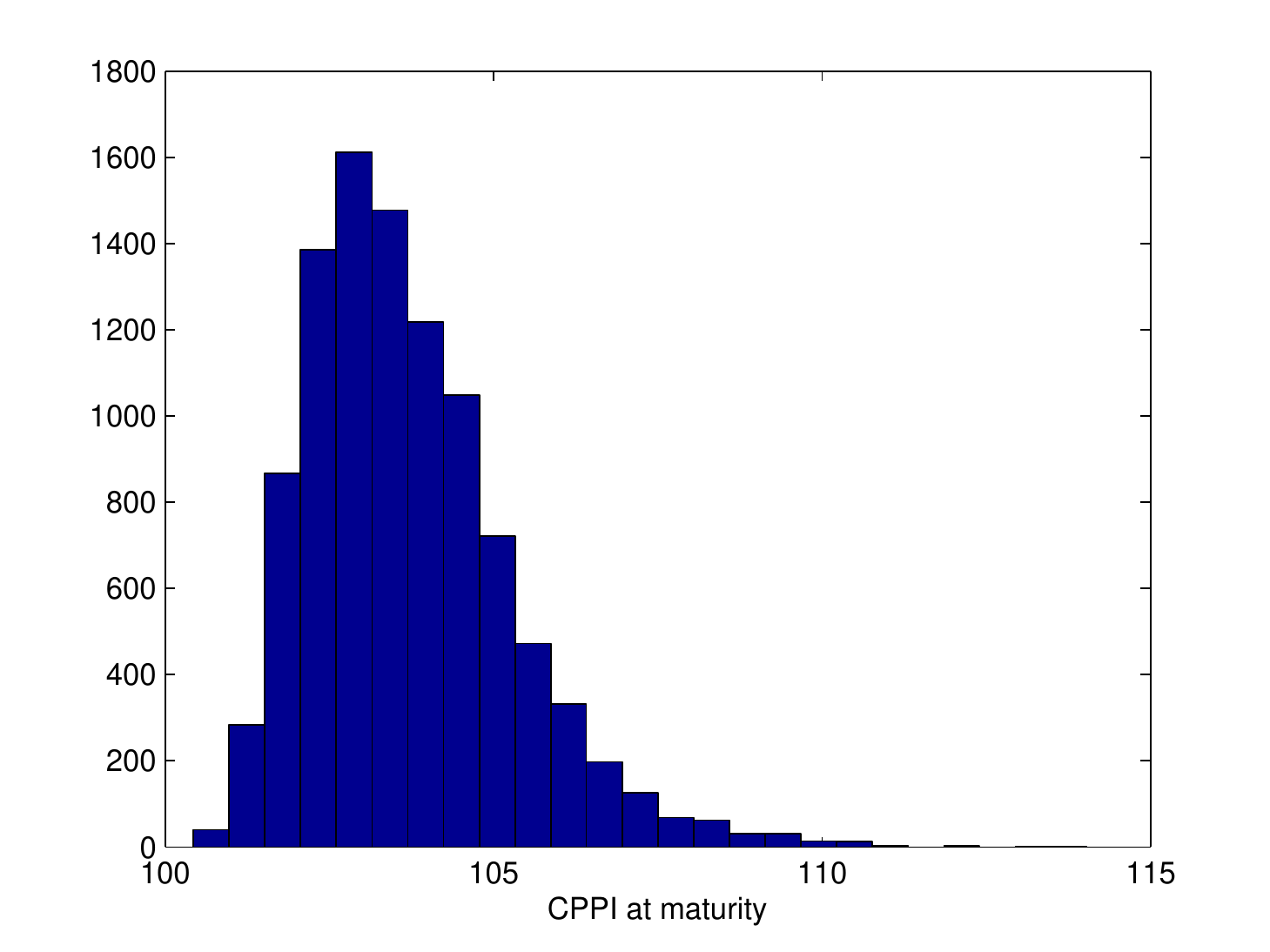}}
        \caption{CPPI-Daily-CL 99\% }
        \raisebox{-\height}{\includegraphics[width=\textwidth]{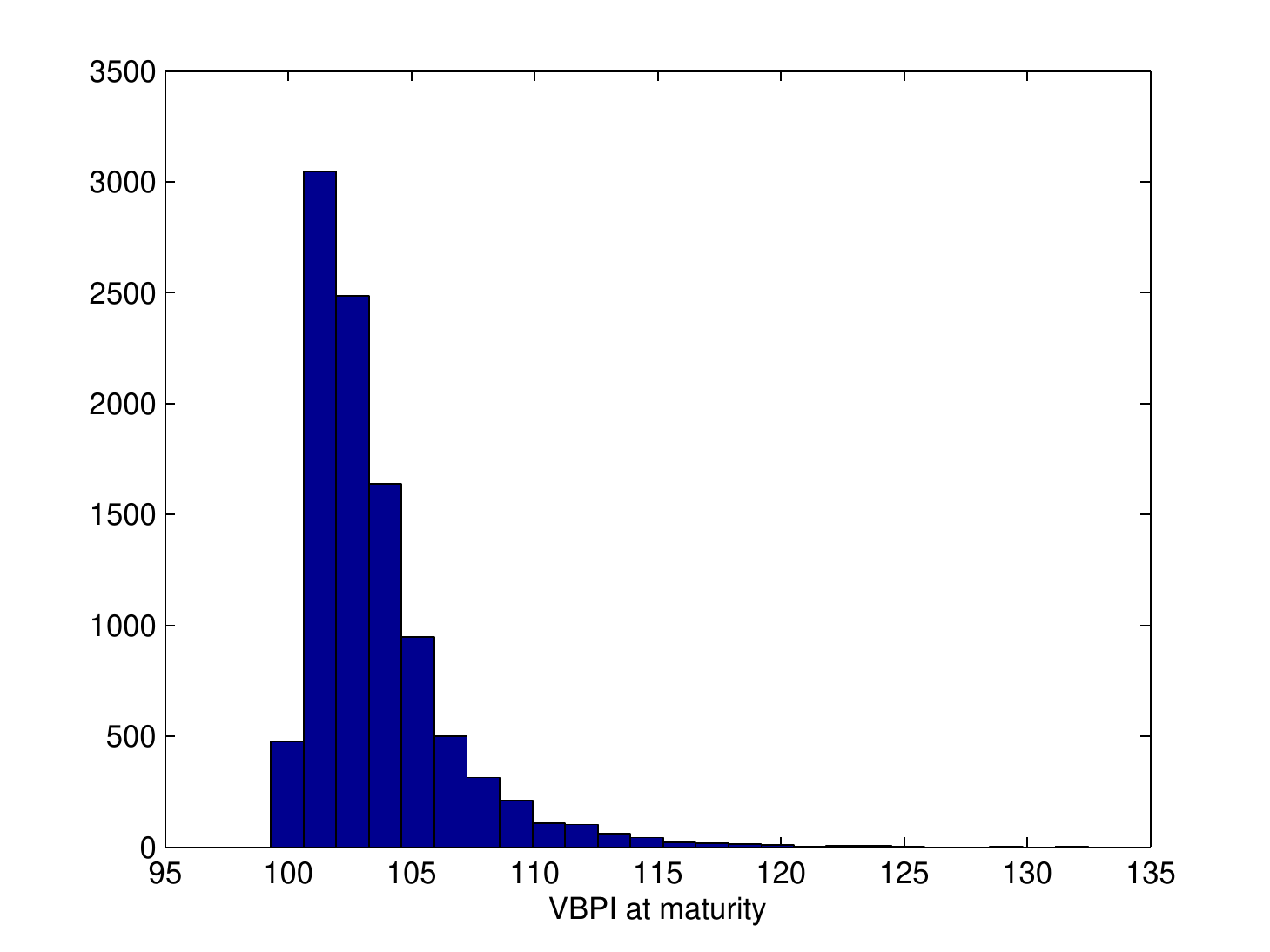}}
        \caption{VBPI-Daily-CL 99\% }
    \end{subfigure}
    \hfill
    \begin{subfigure}[t]{0.32\textwidth}
        \raisebox{-\height}{\includegraphics[width=\textwidth]{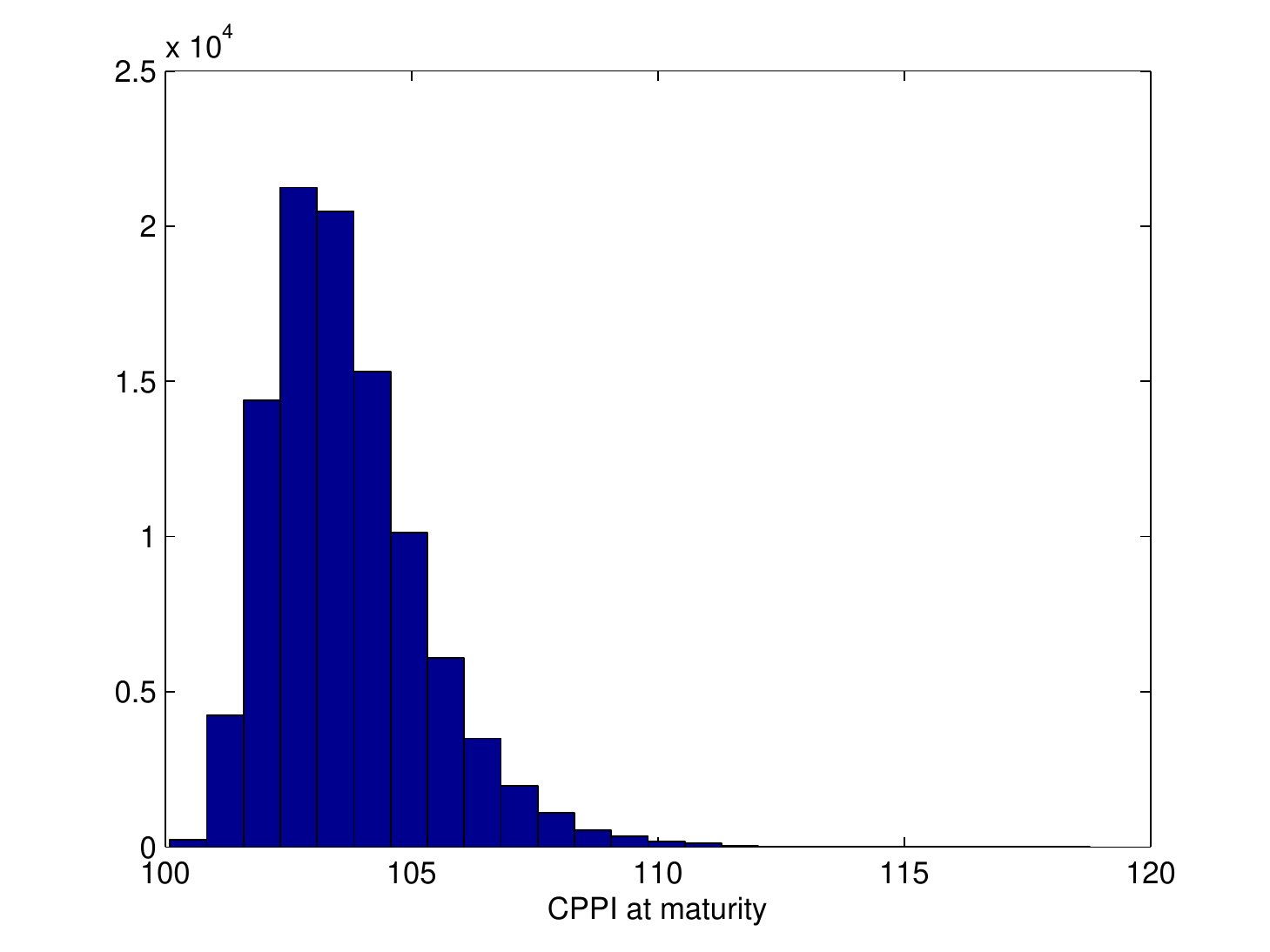}}
        \caption{CPPI-Weekly-CL 99\% }
        \raisebox{-\height}{\includegraphics[width=\textwidth]{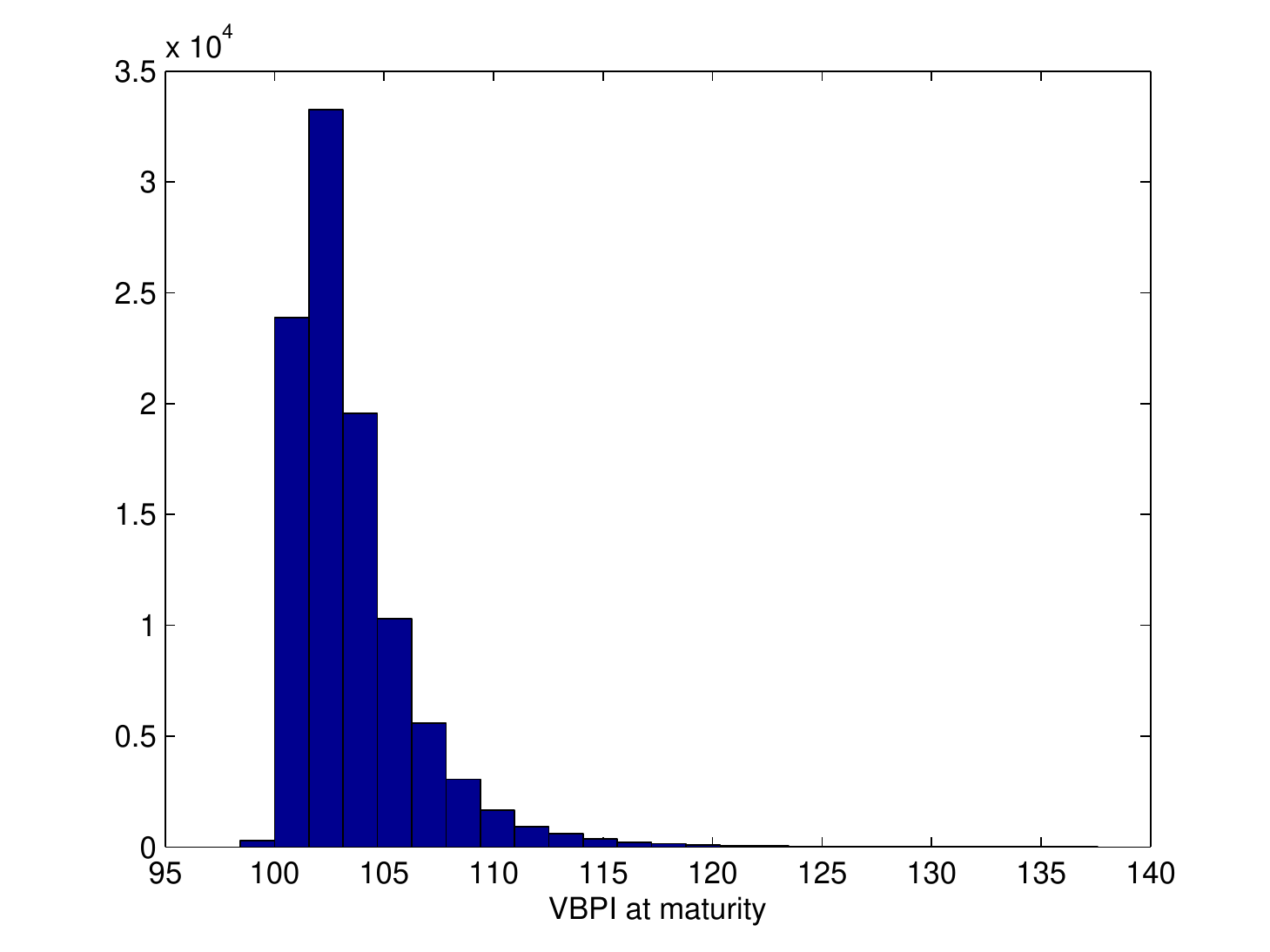}}
        \caption{VBPI-Weekly-CL 99\% }
    \end{subfigure}
    \hfill
    \begin{subfigure}[t]{0.32\textwidth}
        \raisebox{-\height}{\includegraphics[width=\textwidth]{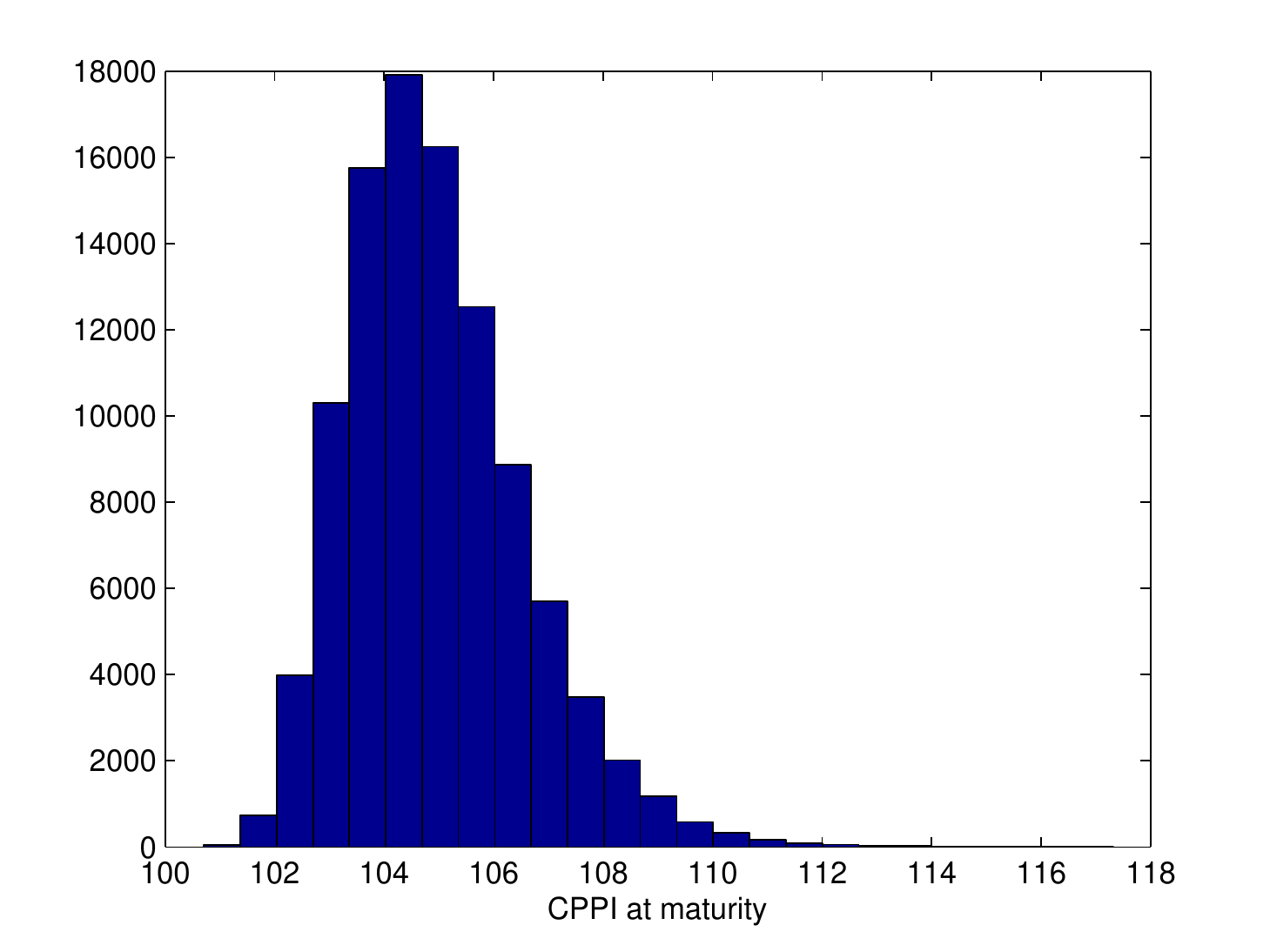}}
        \caption{CPPI-Monthly-CL 99\% }
        \raisebox{-\height}{\includegraphics[width=\textwidth]{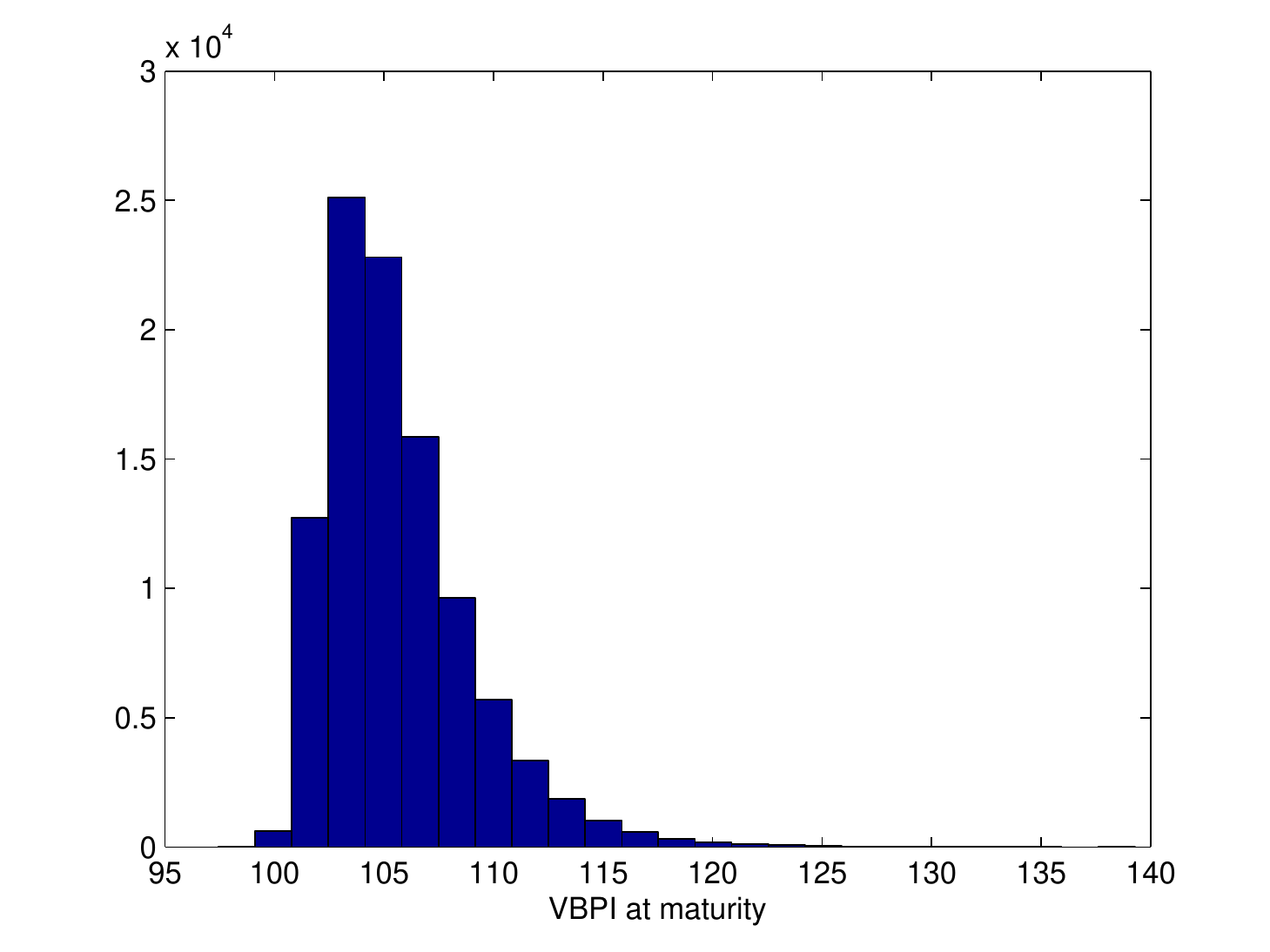}}
        \caption{VBPI-Monthly-CL 99\% }
    \end{subfigure}
        \caption{Frequency of the terminal value of the CPPI and VBPI portfolios at $99\%$ CL}
\end{figure}

\begin{figure}
     \centering
    \begin{subfigure}[t]{0.45\textwidth}
        \raisebox{-\height}{\includegraphics[width=\textwidth]{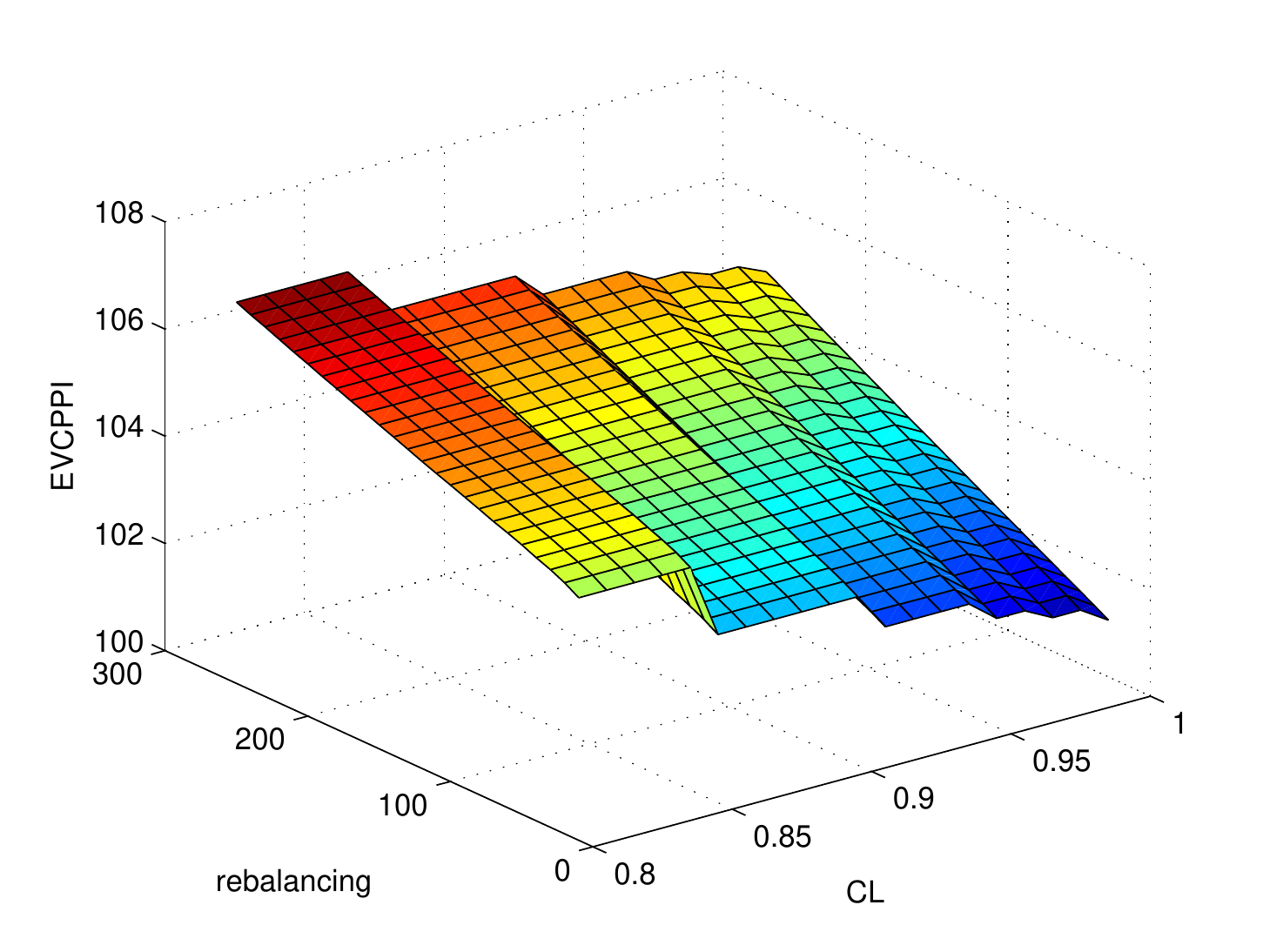}}
        \caption{Expected value of CPPI portfolio. }
    \end{subfigure}
    \hfill
    \begin{subfigure}[t]{0.45\textwidth}
        \raisebox{-\height}{\includegraphics[width=\textwidth]{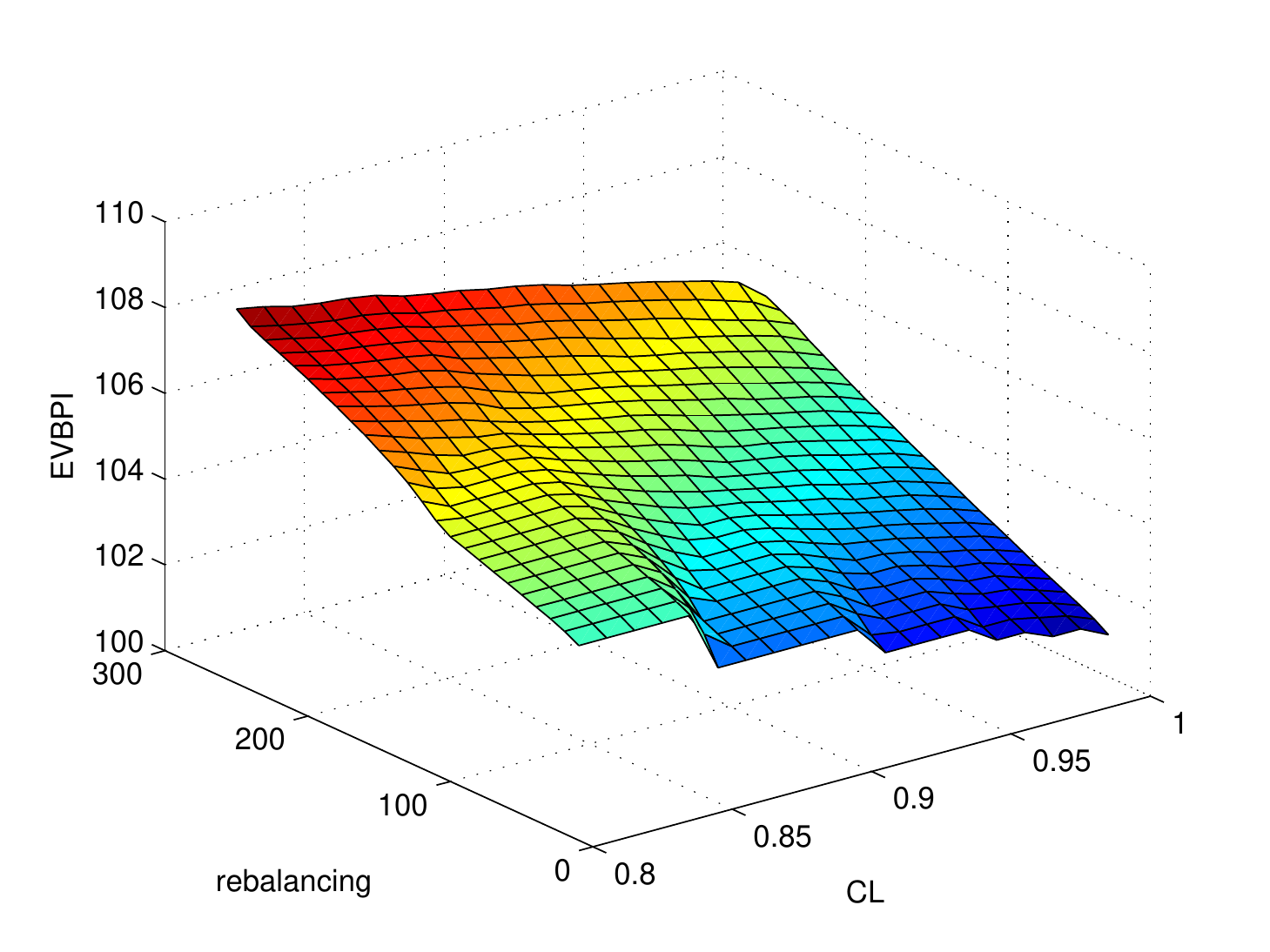}}
        \caption{Expected value of VBPI portfolio.}
    \end{subfigure}
    \caption{Comparing terminal expected value of the CPPI and VBPI portfolios.}
\end{figure}
\begin{figure}
     \centering
    \begin{subfigure}[t]{0.45\textwidth}
        \raisebox{-\height}{\includegraphics[width=\textwidth]{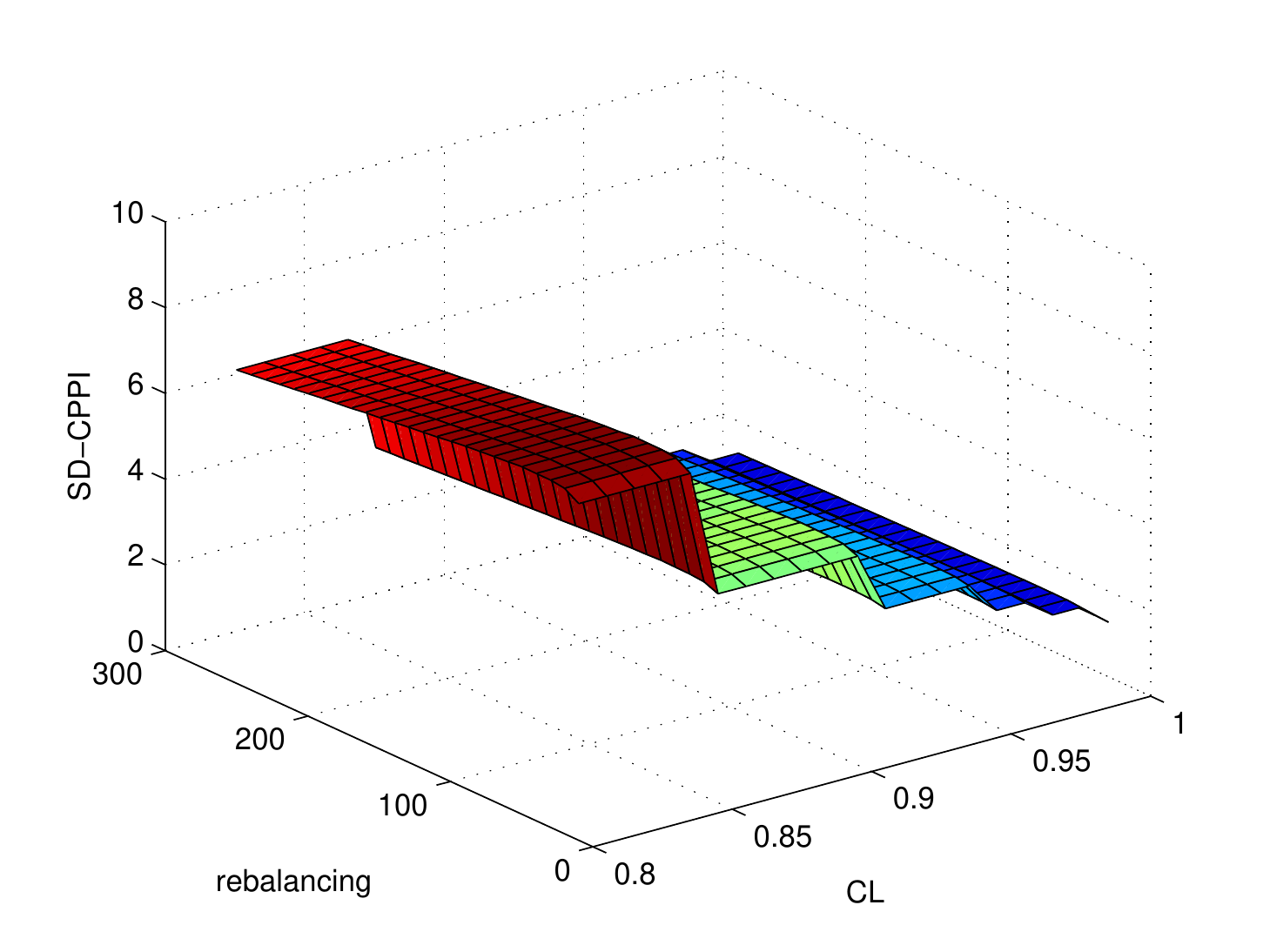}}
        \caption{Standard deviation of the CPPI portfolio. }
    \end{subfigure}
    \hfill
    \begin{subfigure}[t]{0.45\textwidth}
        \raisebox{-\height}{\includegraphics[width=\textwidth]{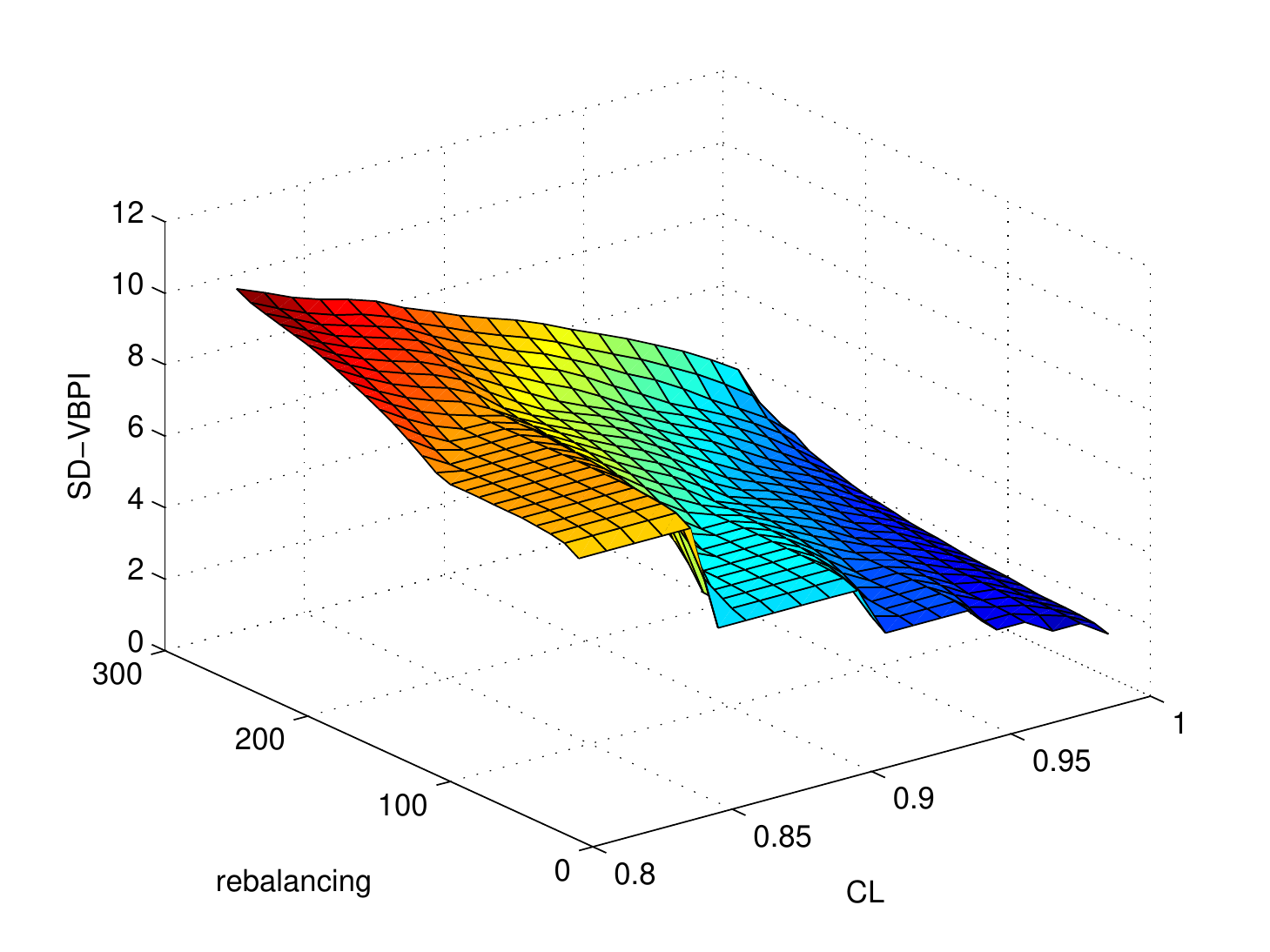}}
        \caption{Standard deviation of the VBPI portfolio.}
    \end{subfigure}
    \caption{Comparing Standard deviation of CPPI and VBPI portfolios.}
\end{figure}
\begin{figure}
     \centering
    \begin{subfigure}[t]{0.45\textwidth}
        \raisebox{-\height}{\includegraphics[width=\textwidth]{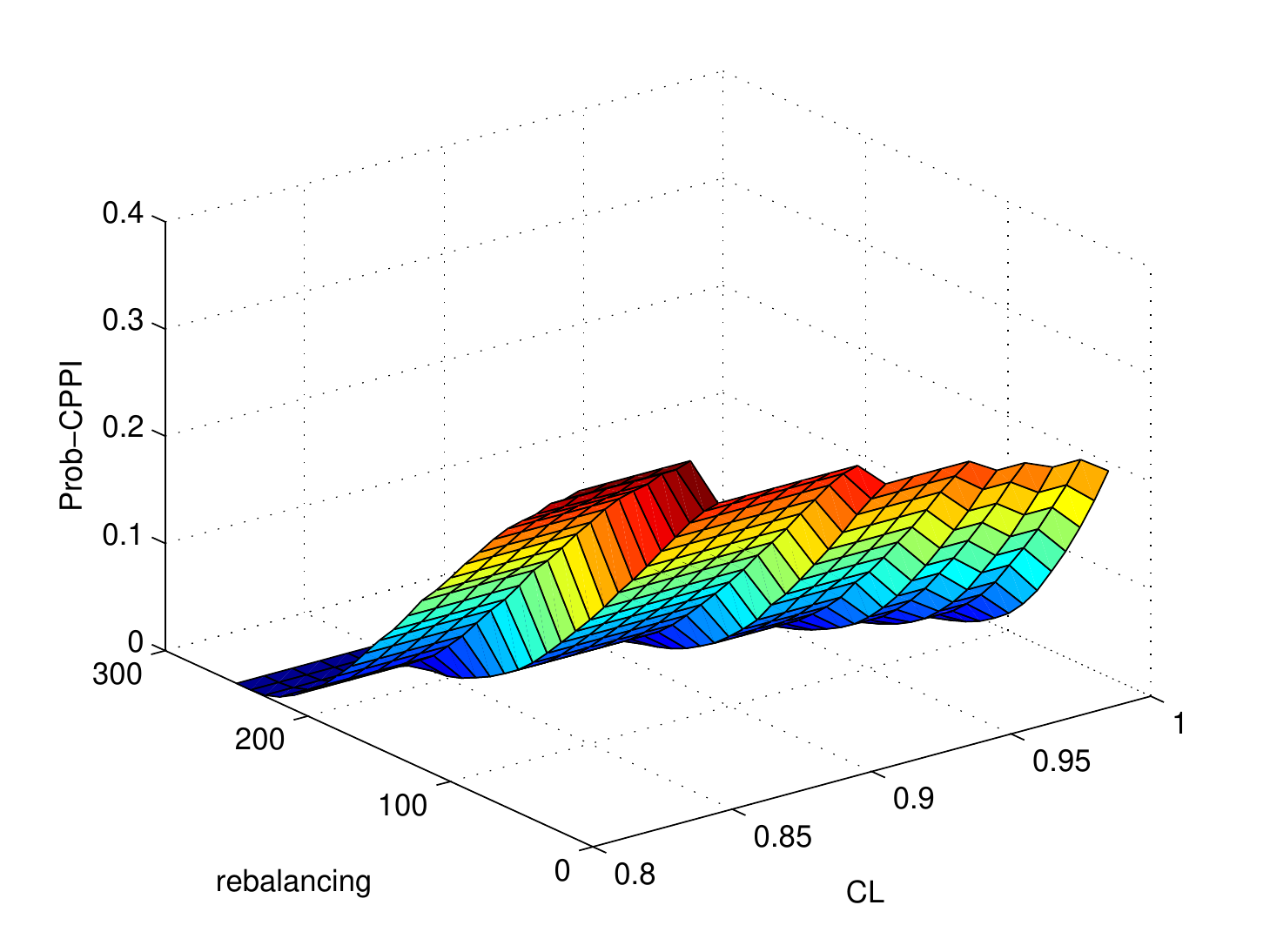}}
        \caption{Shortfall probability of the CPPI portfolio. }
    \end{subfigure}
    \hfill
    \begin{subfigure}[t]{0.45\textwidth}
        \raisebox{-\height}{\includegraphics[width=\textwidth]{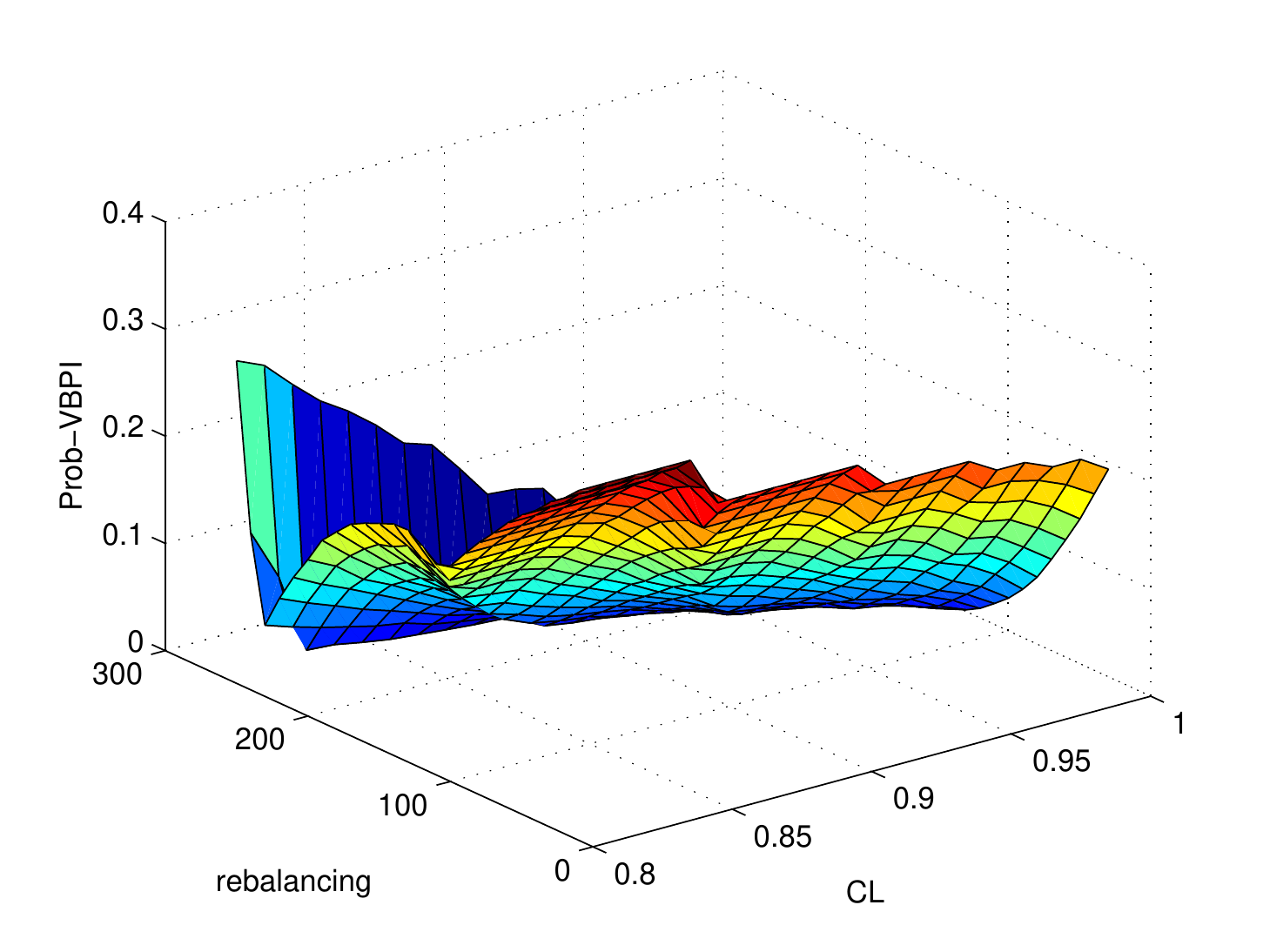}}
        \caption{Shortfall probability of the VBPI portfolio.}
    \end{subfigure}
    \caption{Comparing shortfall probability of the CPPI and VBPI portfolios.}
\end{figure}

\begin{figure}
     \centering
    \begin{subfigure}[t]{0.45\textwidth}
        \raisebox{-\height}{\includegraphics[width=\textwidth]{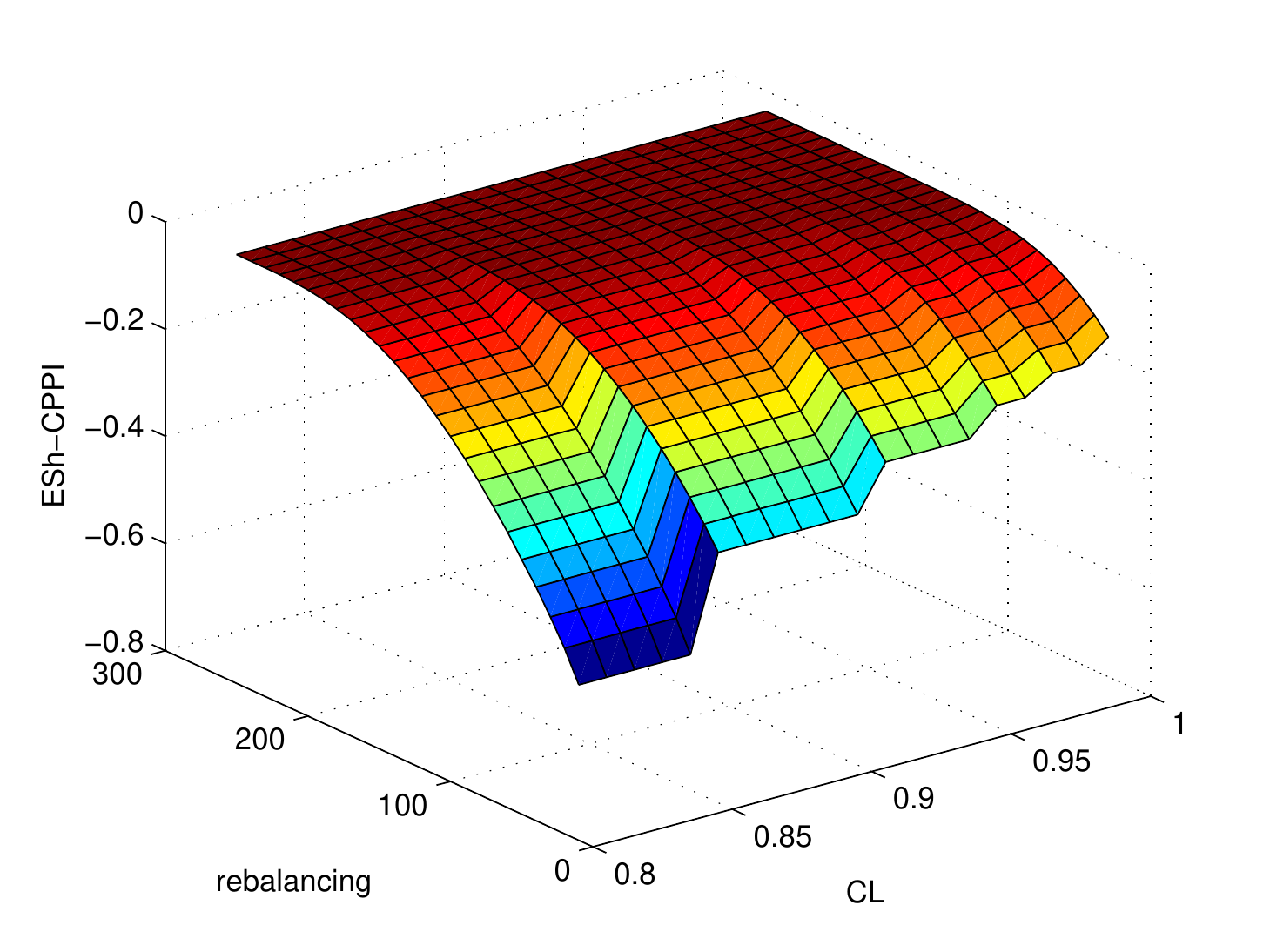}}
        \caption{Expected shortfall of the CPPI portfolio. }
    \end{subfigure}
    \hfill
    \begin{subfigure}[t]{0.45\textwidth}
        \raisebox{-\height}{\includegraphics[width=\textwidth]{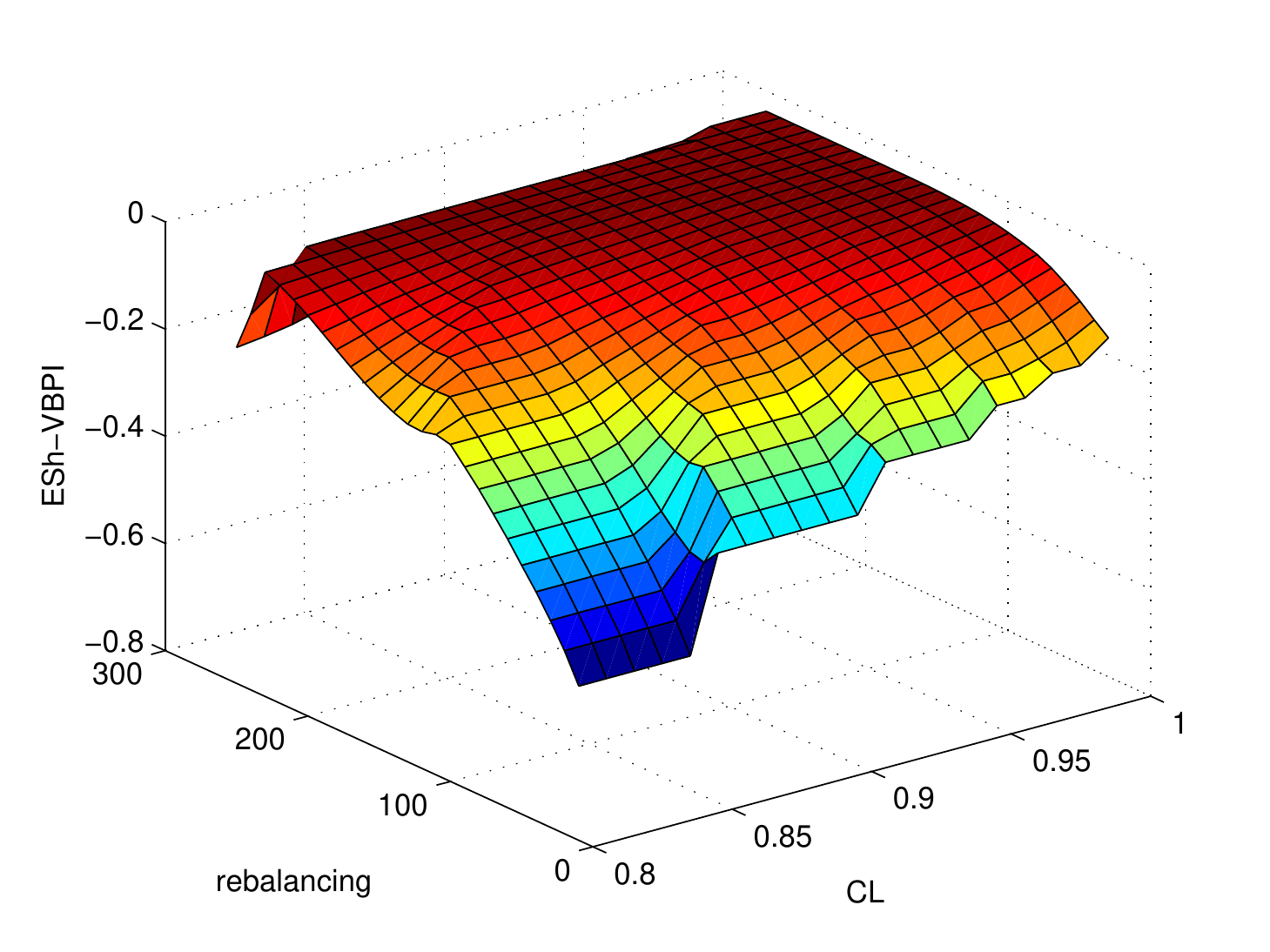}}
        \caption{Expected shortfall of the VBPI portfolio.}
    \end{subfigure}
    \caption{Comparing expected shortfall of the CPPI and VBPI portfolios.}
\end{figure}

\begin{table}[]
\resizebox{\textwidth}{!}{%
\begin{tabular}{ccclllllllllllllll}
\hline
   &               & \multicolumn{3}{c}{Omega CPPI} &  & \multicolumn{3}{c}{Omega VBPI} &  &  & \multicolumn{3}{c}{Kappa CPPI} &  & \multicolumn{3}{c}{Kappa VBPI} \\ \cline{3-5} \cline{7-9} \cline{12-14} \cline{16-18}
CL &               & Daily   & Weekly  & Monthly    &  & Daily   & Weekly   & Monthly   &  &  & Daily   & Weekly   & Monthly   &  & Daily   & Weekly   & Monthly   \\ \hline
   & Thershold 1\% &         &         &            &  &         &          &           &  &  &         &          &           &  &         &          &           \\
90 &               & 46.48   & 0.71    & 0.49       &  & 4.08    & 0.72     & 0.49      &  &  & 15.61   & -0.22    & -0.41     &  & 1.81    & -0.22    & -0.41 \\
95 &               & 629.76  & 0.66    & 0.37       &  & 7.66    & 0.67     & 0.37      &  &  & 85.65   & -0.25    & -0.49     &  & 2.97    & -0.24    & -0.49 \\
99 &               & 947.24  & 0.61    & 0.27       &  & 99.87   & 0.62     & 0.27      &  &  & 425.12  & -0.28    & -0.57     &  & 25.62   & -0.27    & -0.56 \\
   & Thershold 2\% &         &         &            &  &         &          &           &  &  &         &          &           &  &         &          &           \\
90 &               & 4.74    & 0.37    & 0.28       &  & 1.91    & 0.38     & 0.28      &  &  & 2.14    & -0.52    & -0.61     &  & 0.60    & -0.51    & -0.60  \\
95 &               & 12.63   & 0.25    & 0.16       &  & 2.65    & 0.29     & 0.16      &  &  & 4.87    & -0.6     & -0.71     &  & 0.99    & -0.58    & -0.71 \\
99 &               & 50.45   & 0.16    & 0.08       &  & 6.75    & 0.19     & 0.09      &  &  & 13.04   & -0.68    & -0.79     &  & 2.96    & -0.66    & -0.78 \\
   & Thershold 3\% &         &         &            &  &         &          &           &  &  &         &          &           &  &         &          &           \\
90 &               & 1.36    & 0.22    & 0.17       &  & 1.04    & 0.22     & 0.17      &  &  & 0.26    & -0.67    & -0.72     &  & 0.03    & -0.67    & -0.72 \\
95 &               & 2.02    & 0.12    & 0.08       &  & 1.21    & 0.14     & 0.08      &  &  & 0.62    & -0.76    & -0.81     &  & 0.15    & -0.74    & -0.81 \\
99 &               & 3.37    & 0.05    & 0.03       &  & 1.65    & 0.07     & 0.03      &  &  & 1.25    & -0.83    & -0.88     &  & 0.44    & -0.82    & -0.87 \\
   & Thershold 4\% &         &         &            &  &         &          &           &  &  &         &          &           &  &         &          &           \\
90 &               & 0.58    & 0.14    & 0.11       &  & 0.63    & 0.14     & 0.11      &  &  & -0.33   & -0.76    & -0.79     &  & -0.29   & -0.76    & -0.79 \\
95 &               & 0.57    & 0.06    & 0.04       &  & 0.66    & 0.08     & 0.04      &  &  & -0.32   & -0.84    & -0.87     &  & -0.26   & -0.82    & -0.87  \\
99 &               & 0.58    & 0.02    & 0.01       &  & 0.62 & 0.03
& 0.02      &  &  & -0.29   & -0.90    & -0.92     &  & -0.29   &
-0.89    & -0.92 \\ \hline
\end{tabular}%
} \caption{Comparing the performance of CPPI and VBPI strategies via the Omega and Kappa performance measures.}
\end{table}

\begin{table}[]
\small \centering
\begin{tabular}{lllllllll}
\hline
   &  & \multicolumn{3}{c}{CPPI} &  & \multicolumn{3}{c}{VBPI} \\ \cline{3-5} \cline{7-9}
CL &  & Daily & Weekly & Monthly &  & Daily & Weekly & Monthly \\
\hline
90 &  & 0.36 & -0.19  & -0.31   &  & 0.36  & -0.18  & -0.31   \\
95 &  & 0.41 & -0.64   & -0.86   &  & 0.37 & -0.47  & -0.86   \\
99 &  & 0.45   & -1.24   & -1.59    &  & 0.41  & -1.06   & -1.53    \\
\hline
\end{tabular}
\caption{Comparing the performance of CPPI and VBPI strategies via the Sharpe performance measure.}
\end{table}

\section{Conclusion}
The main contribution of this paper is the development of a constrained CPPI as well as a VBPI strategy under a regime-switching diffusion model.
In this respect and for the CPPI strategy, we derive a stochastic differential equation for the portfolio value consisting of a risky and a riskless asset where the exposure is constrained both from below and above.
In the VBPI case, we derive expressions for the weights of the risky and riskless assets in the portfolio based on the value at risk of the portfolio return process at each discrete rebalancing time. We employ a Fourier-based method to approximate the VaR by inverting the characteristic function of the underlying asset. We compare the two approaches based on some performance measures and show that the constrained CPPI method performs well in most of the scenarios examined and provides us with a better control on the gap risk of the investment strategy.
For future research, we propose to add a jump term into the model to better capture the real effects present in the market environment.
We could also consider a flexible floor value which potentially improves the efficiency of the proposed portfolio insurance strategies.

\appendix
\section{Monte-Carlo Simulation Method}
In the Monte-Carlo approach, we first subdivide the time interval $[0,T]$ into $N = \frac{T}{\triangle t}$
 grid points and then simulate a number, $M$, of risky asset sample paths at these discrete points denoted as
$\{S(t_k), k = 1, . . . , N\}$. To do so, it suffices to specify the
values $ {\alpha(t_k), k = 1, . . . , N}$ and then simulate
$V(t_k)$. We use the following method from Yuan
and Mao \cite{mao2006euler} to obtain a sample realization of ${\alpha(t), 0\leq t\leq T}$:
\begin{itemize}
\item[1] Let $\alpha(t_0)=\alpha_0$ as the initial value of $\alpha(t)$;
\item[2] Generate $u\thicksim U([0,1])$ from a uniform distribution in the
interval (0,1). Let $\alpha(t_k-1) = i$ for some $i \epsilon
\mathcal{H} $ be the governed regime at time $t_{k-1}$ . Now define
\begin{equation}\label{vtc}
   \alpha(t_n) =  \left\{
\begin{array}{ll}
    \alpha_n, \ \ \  where \ \ \alpha_n \varepsilon \mathcal{H} \backslash \{H\} \ \ \ if   \ \ \ \  \Sigma ^{\alpha_{n - 1}} _ {j=1} P^{\Delta t}_{i,j}\leq u < \Sigma ^{\alpha_{n}} _ {j=1} P^{\Delta t}_{i,j}\\
    H,     \ \ \ \ \ \ \ \ \ \ \ \ \ \ \ \ \ \ \ \ \ \ \ \ \ \ \ \ \ \ \ \  if  \ \ \  \Sigma^ {H-1} _{j=1} P^{\Delta t}_{i,j}\leq u\\
\end{array}%
\right.
\end{equation}
in which $\Sigma^ {0} _{j=1} P^{\Delta t}_{i,j} = 0$.
\end{itemize}
Iterating this algorithm, we obtain the values $\{\alpha(1),\cdots,\alpha(t_N)\}$ and then we could simulate $V_t^{\rm CPPI}$ (and also $V_t^{\rm VBPI}$) using the following expression:
\begin{equation}\label{vtc}
   {V_{t_{n+1}}} = {V_{t_{n}}} +  \left\{
\begin{array}{ll}
    \displaystyle{{r V_{t_{n}}} \Delta t} , & \hbox{$C_{t_n}\leqslant0$,} \\
    \displaystyle{C_{t_n} (m(\mu_{\alpha_{t_n}}-r)+r)\Delta t + m \sigma_{\alpha_{t_n}} \Delta W_{t_n}}, & \hbox{$0<m C_{t_n}<p {V_{t_n}}$,} \\
    \displaystyle{C_{t_n} (p(\mu_{\alpha_{t_n}}-r)+r)\Delta t + p \sigma_{\alpha_{t_n}} \Delta W_{t_n}}, & \hbox{$p{V_{t_n}} \leqslant m C_{t_n}$.} \\
\end{array}%
\right.
\end{equation}

\end{document}